\documentclass[%
superscriptaddress,
 amsmath,amssymb,
 aps,
 prm,
 twocolumn,
]{revtex4-2}

\usepackage{graphicx}
\usepackage{dcolumn}
\usepackage{bm}
\usepackage{float}
\usepackage{xcolor}
\usepackage{appendix}
\usepackage{multirow}
\usepackage{makecell}
\usepackage{comment}
\usepackage{soul}

\begin{document}

\title{Characterizing the mechanical response of metallic glasses to uniaxial tension using a spring network model}
\author{Aya Nawano}
\affiliation{Department of Mechanical Engineering and Materials Science, Yale University, New Haven, Connecticut 06520, USA}
\author{Jan Schroers}
\affiliation{Department of Mechanical Engineering and Materials Science, Yale University, New Haven, Connecticut 06520, USA}
\author{Mark D.\ Shattuck}
\affiliation{Benjamin Levich Institute and Physics Department, The City College of New York, New York, New York 10031, USA}
\author{Corey S.\ O'Hern}
 \email{corey.ohern@yale.edu}
\affiliation{Department of Mechanical Engineering and Materials Science, Yale University, New Haven, Connecticut 06520, USA}
\affiliation{Department of Physics, Yale University, New Haven, Connecticut 06520, USA}
\affiliation{Department of Applied Physics, Yale University, New Haven, Connecticut 06520, USA}
\affiliation{Graduate Program in Computational Biology and Bioinformatics, Yale University, New Haven, Connecticut 06520, USA}

\begin{abstract}
Metallic glasses are frequently used as structural materials.  Therefore, it is important to develop methods to predict their mechanical response as a function of the microstructure prior to loading. We develop a novel coarse-grained spring network model, which describes the mechanical response of metallic glasses using an equivalent series network of springs, which can break and re-form to mimic atomic rearrangements during deformation. To validate the spring network model, we perform numerical simulations of quasistatic, uniaxial tensile deformation of Lennard-Jones and embedded atom method (EAM) potentials for Cu$_{50}$Zr$_{50}$ metallic glasses in the absence of large-scale shear band formation. We consider samples prepared using a wide range of cooling rates and with different amounts of crystalline order. We show that both the Lennard-Jones and EAM models possess qualitatively similar stress $\sigma$ versus strain $\gamma$ curves. By specifying five parameters in the spring network model (ultimate strength, strain at ultimate strength, slopes of $\sigma(\gamma)$ at $\gamma=0$ and at large strain, and strain at fracture where $\sigma=0$), we can accurately describe the form of the stress-strain curves during uniaxial tension for the computational studies of Cu$_{50}$Zr$_{50}$, as well as recent experimental studies of several Zr-based metallic glasses. For the computational studies of Cu$_{50}$Zr$_{50}$, we find that the yield strain distribution is shifted to larger strains for slowly cooled glasses compared to rapidly cooled glasses. In addition, the average number of new springs and their rate of formation decreases with decreasing cooling rate. These effects offset each other at large strains, causing the stress-strain curve to become independent of the sample preparation protocol in this regime. In future studies, we will extract the parameters that define the spring network model directly from atomic rearrangements that occur during uniaxial deformation. 

\end{abstract}
\maketitle

\section{INTRODUCTION}

\begin{figure*}
\centering
\includegraphics[scale=0.8]{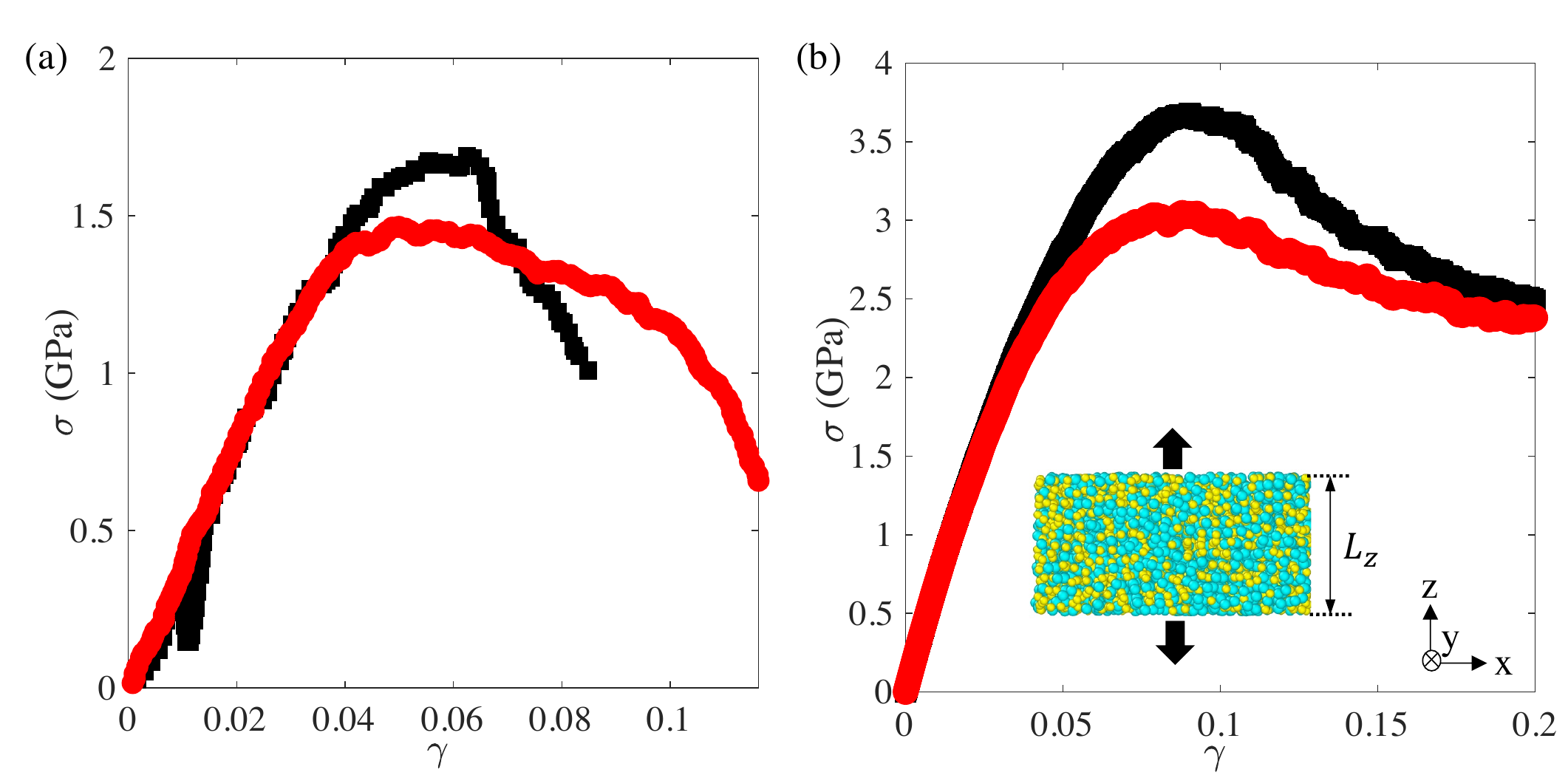}
\caption{Engineering stress $\sigma$ plotted versus engineering strain $\gamma$ from (a) experiments and (b) simulations of metallic glasses undergoing uniaxial tension.  The data in (a) is from sputtered $\rm Zr_{56}Ni_{22}Al_{22}$ metallic glasses~\cite{RN25} with (black) and without (red) annealing at temperatures below the glass transition for $24$ hours. The data in (b) is from athermal, quasistatic uniaxial tension simulations of $\rm Cu_{50}Zr_{50}$ modeled using EAM interactions generated at cooling rates $R = 10^{10} \rm K/s$ (black) and $10^{13} \rm K/s$ (red) and averaged over $50$ samples. The inset shows the geometry used in the simulations. The samples have periodic boundaries in the $z$-direction that are moved vertically to apply tensile deformations and open boundaries in the $x$- and $y$-directions.}
\label{fig1}
\end{figure*}

Bulk metallic glasses are alloys with amorphous atomic structure. Since they possess larger values for the strength and elastic limit compared to those for conventional crystalline alloys, they represent a promising class of structural materials~\cite{RN65}. However, under tensile loading, metallic glasses at room temperature are typically brittle. Shear bands, or localized regions of large strain, form during deformation that can lead to failure of the material~\cite{RN66}. As an example of this behavior, in Fig.~\ref{fig1} (a), we show the engineering stress $\sigma$ versus engineering strain $\gamma$ from recent experiments that perform uniaxial tension tests on sputtered ZrNiAl metallic glasses~\cite{RN25}. The more ductile sample has an ultimate strength of $\sigma_m \sim 1.5\rm GPa$ and fractures at $\gamma_f \sim 11.6\%$, whereas the more brittle sample has an ultimate strength of $\sigma_m \sim 1.7\rm GPa$ and fractures at $\gamma_f \sim 8\%$. The two samples were fabricated using similar processes, but the more brittle sample was annealed at a temperature below the glass transition temperature $T_g$ for $24$ hours after fabrication. Many factors have been shown to affect the tensile plasticity of metallic glasses, such as the cooling history\cite{RN109}, sample size~\cite{RN15,RN25,RN12,RN24}, strain rate~\cite{RN9,RN10,RN111}, and temperature~\cite{RN10,RN111} at which the testing occurs. In general, larger samples, samples prepared at lower cooling rates, and samples tested at lower temperatures are more brittle. In addition, experiments on bending and compression of metallic glasses have shown that their mechanical response is influenced by changes in chemical composition~\cite{RN67,RN68,RN69}. For example, adding 5\% of Al atoms into $\rm Cu_{50}Zr_{50}$ bulk metallic glasses increased the failure strain from 7.9\% to 18\% in compression tests~\cite{RN69}. Also, Pd-based bulk metallic glasses were found to be brittle when formed at low cooling rates, whereas Pt-based bulk metallic glasses are ductile regardless of the cooling rate used to prepare the samples~\cite{RN67}. Because there are so many factors that influence the microstructure of metallic glasses, it is difficult to predict the mechanical response of metallic glasses to applied deformations. 

The mechanical response of metallic glasses is controlled by the atomic interactions and motions that arise from applied deformations. Numerous molecular dynamics (MD) simulation studies have shown that the mechanical response of metallic glasses to applied deformations involves highly collective and non-affine atomic motions that are spatially and temporally correlated.  To understand the mechanical response of metallic glasses at larger length scales, coarse-grained mesoscale models have been developed~\cite{RN8,RN2,RN1}. For example, elastoplastic models consider metallic glasses as a collection of mesoscopic elements, where each element deforms elastically until it reaches its local yield strain. After yielding, the elements re-distribute their stress to neighboring elements and the stress is reset to the largest value in the elastic state. Further, elaborations of elastoplastic models have coupled the shape of the yield strain distributions to the evolving microstructure during applied deformation~\cite{RN81}. Several studies have shown that elastoplastic models can describe the stress-strain curves for model Lennard-Jones (LJ) glasses undergoing simple shear over a range of strain rates~\cite{RN34,RN77,RN81}. 

Theoretical models for the irreversible atomic motions that occur during applied deformation can be used to improve elasto-plastic descriptions of metallic glasses. For example, the plastic strain can be calculated by identifying shear transformation zones (STZs), or groups of atoms that undergo collective, non-affine motion in response to applied deformations~\cite{RN90}. Manning, \emph{et al.}~\cite{RN79} derived a system of ordinary differential equations (ODEs) for the deviatoric stress and an effective temperature that controls the STZ density in computational studies of LJ glasses undergoing simple shear. The system of ODEs includes seven parameters, such as the initial and steady-state effective temperature, characteristic size of an STZ, and effective temperature diffusivity, which are chosen so that the predicted stress versus strain and degree of strain localization match the behavior in the numerical simulations. Similar studies have coupled elasto-plastic and STZ descriptions to describe the stress versus strain in molecular dynamics simulations of embedded atom method potentials for Cu$_{50}$Zr$_{50}$ undergoing simple shear~\cite{RN33}.

The fiber bundle model~\cite{RN4,RN5,RN6} was originally developed to describe fibrous materials under tension, but it has also been used to describe amorphous solids undergoing tensile loading~\cite{RN72,RN73}. The fiber bundle model is a coarse-grained, one-dimensional model that considers fibers in parallel under a constant load. An individual fiber breaks when its extension exceeds a randomly selected threshold, and its load is then redistributed to neighboring fibers. This model displays brittle, quasi-brittle, and ductile failure modes as a function of the heterogeneity in the failure thresholds and the length scale over which the stress is redistributed after local failure~\cite{RN72}. Key differences between the fiber bundle and elastoplastic models are that each fiber only experiences elastic deformation before yielding and there is no stress recovery within a fiber after it yields. 

In this article, we develop a spring network model to describe the mechanical response of metallic glasses undergoing tensile loading. The spring network model includes a large number $N_s$ of initially unstretched springs in parallel prior to the applied deformation; these springs stretch during a series of small applied strain steps. Similar to the fiber bundle model, the springs break when their extension exceeds a threshold. However, unlike the conventional fiber bundle model, at each strain step, new springs form, contributing to the stress that resists the tensile load. We implement constant strain instead of constant stress boundary conditions in the spring network model to compare to molecular dynamics simulations of metallic glasses undergoing tensile loading performed at constant strain.

In the $N_s \rightarrow \infty$ limit, we derive an analytical form for the stress $\sigma$ versus strain $\gamma$ for the spring netowrk model undergoing tensile loading. This expression includes five important parameters: the ultimate strength, strain at which this occurs, slopes of $\sigma(\gamma)$ at zero and at large strain, and failure strain at which $\sigma=0$. This expression for $\sigma(\gamma)$ accurately describes the mechanical response of both Lennard-Jones (LJ) and embeded atom method (EAM) models for $\rm Cu_{50}Zr_{50}$ metallic glasses prepared over a range of cooling rates, possessing a range of local crystalline order, and evolving in the absence of large-scale shear band formation.  We chose to study $\rm Cu_{50}Zr_{50}$ in the numerical simulations because it is one of the few binary alloys that forms bulk metallic glasses~\cite{RN57}; experimental studies have shown that $\rm Cu_{50}Zr_{50}$ has a critical cooling rate of $R_c \sim250 \rm K/s$ \cite{RN42}. By comparing the computational results with predictions from the spring network model, we show that the yield strain distribution is shifted to larger strains for slowly cooled compared to rapidly cooled glasses. In addition, the average number of new springs and their rate of formation decreases with decreasing cooling rate. These effects offset each other at large strains, causing the stress-strain curve to become independent of sample preparation protocol in this regime. We also show that $\sigma(\gamma)$ obtained from the spring network model accurately describes the mechanical response of several Zr-based metallic glasses, including $\rm Zr_{65}Al_{10}Ni_{10}Cu_{15}$, Zr$_{56}$Ni$_{22}$Al$_{22}$, and Cu$_{49}$Zr$_{51}$, obtained in recent experimental studies. In the current work, we relate the macroscopic mechanical response of metallic glasses to mesoscopic spring elements in the coarse-grained spring network model. In future work, we will define the spring elements in terms of atomic rearrangements that occur during uniaxial loading in all-atom MD simulations. This approach will allow us to better understand how non-affine atomic motions determine the macroscopic mechanical behavior of amorphous materials in the absence of large-scale shear banding.  

The remainder of the article is organized as follows. In Sec.~\ref{methods}, we first describe the numerical simulations of Cu$_{50}$Zr$_{50}$ metallic glasses (modeled using the LJ and EAM potentials) undergoing athermal, quasistaitc uniaxial tension. We then introduce the spring network model and derive an analytical expression for stress versus strain during tensile loading. In Sec.~\ref{results}, we present the results. We first show the maximum stress during tensile loading as a function of the total potential energy of the undeformed structure. We then show that the expression for $\sigma(\gamma)$ from the spring network model accurately describes the mechanical response during tensile loading obtained from simulations (for both LJ and EAM models) and recent experiments on Zr-based metallic glasses. In Sec.~\ref{conclusions}, we put forward our conclusions and discuss future work on extracting the spring network parameters directly from atomistic simulations. Finally, in Appendix ~\ref{appendix}, we discuss the effects of system size on the mechanical response from simulations of athermal, quasistatic uniaxial tension. 

\section{Methods}
\label{methods}

In this section, we first describe the interaction potentials that are used to model $\rm Cu_{50}Zr_{50}$ metallic glasses. We then introduce the cooling and structural relaxation protocols to prepare $\rm Cu_{50}Zr_{50}$ metallic glasses with different amounts of local positional order. We also describe the simulation method for applying the tensile deformation to the metallic glass samples. We then present the spring network model, derive an analytical expression for the stress versus strain, $\sigma(\gamma)$, and relate the five parameters in the spring network model to important features of $\sigma(\gamma)$. 

\subsection{Atomic Interaction Potentials for $\rm Cu_{50}Zr_{50}$}

We consider the LJ ~\cite{RN91} and EAM interaction potentials~\cite{RN13} for modeling $\rm Cu_{50}Zr_{50}$ metallic glasses. The LJ potential includes isotropic, pairwise atomic interactions with parameters that control the atomic size and the attractive strength of the interactions. EAM potentials include both pairwise atomic interactions, as well as many-body interactions that arise from the electronic degrees of freedom. 

For the systems with LJ interactions, we employ a truncated and force-shifted potential energy:
\begin{equation}
 \left.U_{ij}(r_{ij})=\phi(r_{ij})-\phi(r_c)-(r_{ij}-r_c)\frac{d\phi_{ij}}{dr_{ij}}\right\vert_{r_{ij}=r_c}
 \label{eq:1}
\end{equation}
for $r_{ij} < r_c$ and $U_{ij}(r_{ij}) = 0$ for $r_{ij} \ge r_c$, where $r_{ij}$ is the separation between atoms $i$ and $j$, $r_c =2.5\sigma_{ij}$, and
\begin{equation}
\phi_{ij}(r_{ij})= 4\epsilon_{ij}\left[(\sigma_{ij}/r_{ij})^{12}-(\sigma_{ij}/r_{ij})^6\right].
\label{LJ}
\end{equation}
The total potential energy is the sum of $U_{ij}(r_{ij})$ over distinct atomic pairs, $U=\sum_{i>j} U_{ij}(r_{ij})$ and the pair force on atom $i$ from $j$ is ${\vec F}_{ij} = (dU/dr_{ij}) {\hat r}_{ij}$, where ${\hat r}_{ij}$ is the unit vector that points from the center of atom $j$ to the center of $i$.

We assume that A-type atoms are Zr and B-type atoms are Cu. We set the the energy and length parameters in Eq.~(\ref{LJ}) as follows:  $\sigma_{\rm BB}/\sigma_{AA}=0.7975$, $\sigma_{\rm AB}/\sigma_{AA}=(1+\sigma_{\rm BB}/\sigma_{AA})/2$, $\epsilon_{\rm BB}/\epsilon_{\rm AA}=0.5584$, and $\epsilon_{\rm AB}/\epsilon_{\rm AA} = (1+\epsilon_{\rm BB}/\epsilon_{\rm AA})/2-\Delta H_{\rm mix}/\epsilon_{\rm AA}=0.8167$ using experimental values of the atomic radii~\cite{RN112}, cohesive energy~\cite{RN53}, and heat of mixing~\cite{RN54}. The mass ratio is set to $m_A/m_B=1.435$, which is the ratio of the molar masses of Zr and Cu. 

Below, when we describe the LJ simulation results, we convert length and energy scales into physical units using $\sigma_{\rm AA}=2.9{\textup{\AA}}$ and $\epsilon_{\rm AA}=0.74\rm eV$~\cite{RN64}.  
When these values are used, Zr atoms on an HCP lattice yield a cohesive energy of $6.47 \rm eV$ and lattice constant of $3.22\textup{\AA}$, which match experimental results~\cite{RN13}. The temperature, pressure, and time scales are then $\epsilon_{\rm AA}/k_B=8.6\times10^3\rm K$, $\epsilon_{\rm AA}/\sigma_{\rm AA}^3=4.7\rm GPa$, and $\sigma_{\rm AA}\sqrt{m_{\rm AA}/\epsilon_{\rm AA}}=0.33\rm ps$, where $k_B$ is Boltzmann's constant.

We also considered an EAM interaction potential to model the mechanical response of Cu$_{50}$Zr$_{50}$ to uniaxial tension. We selected the EAM potential developed by Mendelev {\it et al.}, who studied vitrification of CuZr alloys~\cite{RN13}. For the EAM, the total potential energy is the sum of two terms~\cite{RN107}:
\begin{equation}
 U= \sum_{i=1}^N {\cal F}_i\left(\sum_{j\neq i}\rho_{ij}(r_{ij})\right) + \sum_{i>j} \phi^p_{ij}(r_{ij}),
 \label{EAM}
\end{equation}
where ${\cal F}_i$ is the many-body embedding function that depends on the electron density of atom $i$ due to all other atoms in the system and $\phi^p_{ij}(r_{ij})$ is the pairwise interaction term. Both the many-body and pairwise terms have a cutoff of $r_c =7.6${\AA} beyond which $U=0$.  This EAM potential was calibrated to match the formation energies of the CuZr equilibrium crystal phases at zero temperature, and the atomic density, mixing enthalpy, and partial pair correlation functions at $1000 {\rm K}$ for $\rm Cu_{46}Zr_{54}$. In addition, Zhang, \emph{et al.} \cite{RN93} used this EAM potential to prepare amorphous Cu$_{50}$Zr$_{50}$ samples using a hybrid Monte Carlo (MC) and molecular dynamics (MD) simulation technique at effective cooling rates as low as 500 $\rm K/s$. The structure factor and shear modulus of the slow quenched Cu$_{50}$Zr$_{50}$ samples obtained from the hybrid MC/MD technique were similar to those obtained experimentally at comparable cooling rates.  We also measured the melting temperature $T_m$ of pure Zr and pure Cu with this EAM potential using the method described by Tang and Harrowell~\cite{RN108}. We found $T_m\sim2110\rm K$ for Zr and $\sim 1356 \rm K$ for Cu for this EAM potential, which are similar to the experimental values of $T_m\sim2128\rm K$ and $\sim 1358 {\rm K}$ for Zr and Cu, respectively.  

\begin{figure}
\centering
\includegraphics[width=0.48\textwidth]{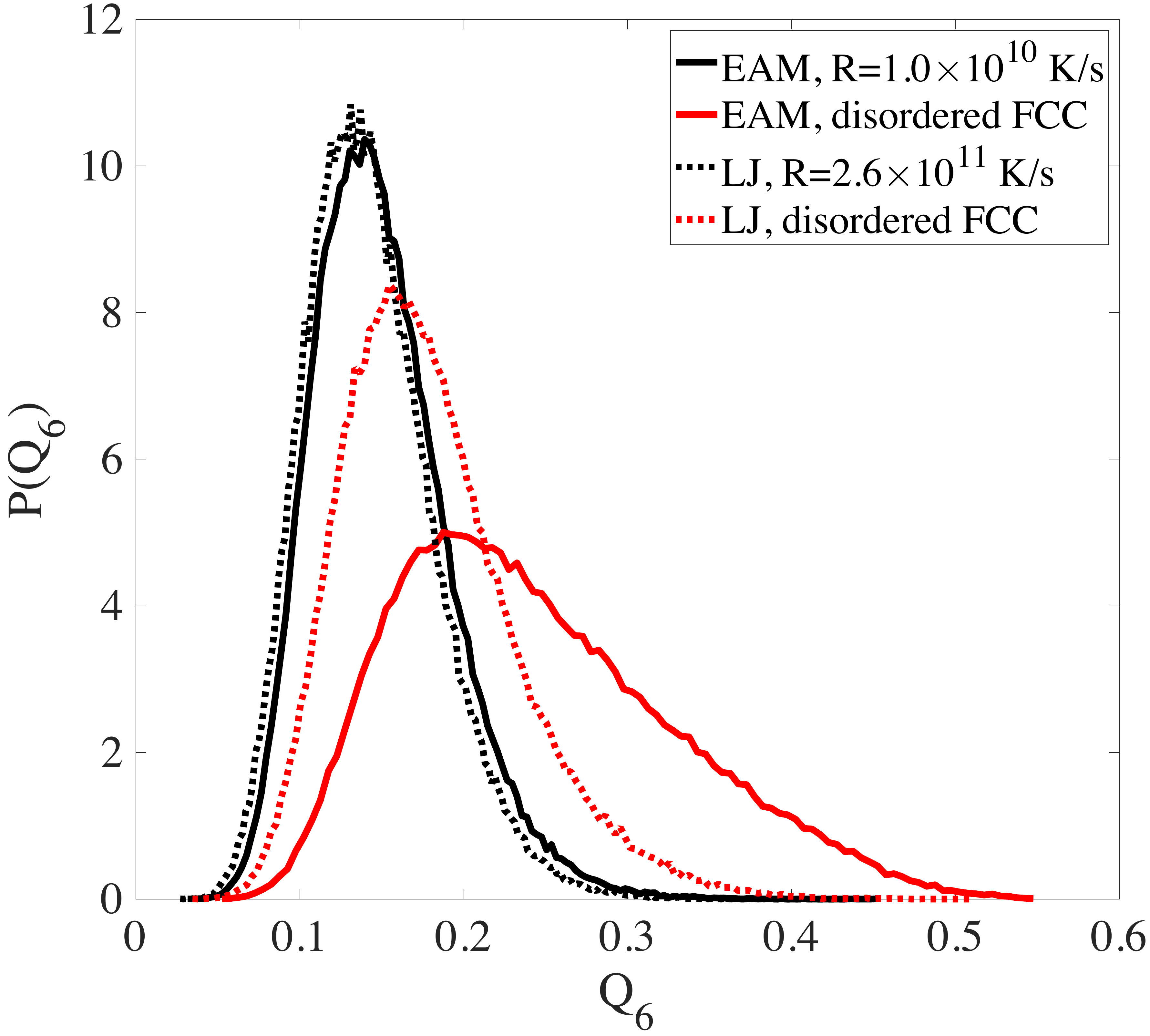}
\caption{Probability distribution of the local bond-orientational order parameter $P(Q_6)$ for each atom in Cu$_{50}$Zr$_{50}$ metallic glasses prepared by thermally quenching at cooling rates, $R=\rm 2.6\times10^{11} \rm K/s$ and $1.0\times10^{10}\rm K/s$ for the LJ and EAM models and in disordered FCC structures obtained by randomly placing Cu and Zr atoms on an FCC lattice followed by potential energy minimization.}
\label{fig2}
\end{figure}

\subsection{Generating Metallic Glasses with Different Amounts of Positional Order}

We focus on systems containing $N=3456$ atoms confined within a cuboidal box with aspect ratios $L_x/L_z = 2$ and $L_y/L_z=2$. When initially preparing the metallic glass samples, we use periodic boundary conditions in the $x$-, $y$-, and $z$-directions. We show in Appendix A that above this system size $\sigma(\gamma)$ is independent of $N$. Numerous studies have shown that the brittleness of metallic glass samples increases with aspect ratio~\cite{RN83}. The values $L_x/L_z = 2$ and $L_y/L_z=2$ give rise to rather ductile response, although slow cooling rates and enhanced positional order can induce more brittle response even for these small aspect ratios.  

To prepare the metallic glass samples, we first equilibrated the systems at high temperature above the melting temperature, $T > T_m$, and then cooled them linearly to low temperature $T_0 < 1 \rm K$ at constant low pressure $P_0$ that is several orders of magnitude below the maxima in the shear stress versus strain curves (obtained from uniaxial tension) using the Nos\'{e}-Hoover thermostat and barostat. The equations of motion are integrated using a modified velocity-verlet algorithm with time step $\rm \Delta t=10^{-3} ps$. The cooling rates spanned four orders of magnitude from $10^{10}$ to $10^{14} \rm K/s$, but remain much larger than the critical cooling rate $R_c \sim 250 {\rm K/s}$ for Cu$_{50}$Zr$_{50}$. After thermally quenching the samples, they were decompressed and potential energy minimized using the conjugate gradient method to reach $P=P_0$ and zero temperature. The maximum total force on an atom after potential energy minimization was $\rm 10^{-10}eV/\textup{\AA}$.  We also performed instantaneous thermal quenches by equilibrating the systems at $T > T_m$ and then minimizing the total potential energy at a constant volume that corresponds to $P=P_0$.

In addition to thermally quenched glasses, we also generated disordered face centered cubic (FCC) structures to span a wider range of brittle and ductile mechanical response to uniaxial tensile deformation. We first placed Zr and Cu atoms randomly on an FCC lattice (while maintaining the correct stoichiometry for Cu$_{50}$Zr$_{50}$) followed by potential energy minimization. The FCC lattice is unstable for random mixtures of Cu and Zr, and thus potential energy minimization induces positional disorder. (Note that pure Cu forms FCC and pure Zr forms hexagonal close packed (HCP) crystalline structures.)  

We can characterize the degree of positional order in a given atomic configuration using the six-fold local bond orientational order (BOO) parameter $Q_6$, which gives the degree of six-foled orientational symmetry of that atom's nearest neighbors~\cite{RN61}. Typically, $Q_6 \gtrsim 0.25$ for a crystal-like atom and $Q_6$ takes on smaller values for atoms in icosahedral or other amorphous structural motifs~\cite{RN85,RN86}. Atoms in FCC and HCP lattices have $Q_6=0.575$ and $0.484$, respectively. 

The $l$-fold local BOO parameter of atom $i$ is defined using
\begin{equation}
q_{lm}(i) = \frac{1}{N_i}\sum_{j=1}^{N_i}\frac{A_{ij}}{A_{tot}^i}Y_{lm}(\theta( \textbf r_{ij}),\phi( \textbf r_{ij})),
\label{eq:2}
\end{equation}
where $N_i$ is the number of Voronoi neighbors~\cite{RN58} of atom $i$, $A_{ij}$ is the area of the Voronoi face shared by atoms $i$ and $j$, $A_{\rm tot}^i$ is the total area of all faces belonging to the Voronoi polyhedron of atom $i$, $Y_{lm}(\theta( {\vec r}_{ij}),\phi( {\vec r}_{ij}))$ is the spherical harmonic function of degree $l$ and order $m$, $\theta({\vec r}_{ij})$ and $\phi({\vec r}_{ij})$ are the polar and azimuthal angles that parameterize the orientation of the vector ${\vec r}_{ij}$ connecting atoms $i$ and $j$~\cite{RN60}. We then average $q_{lm}(i)$ to obtain \cite{RN59}:
\begin{equation}
Q_{lm}(i) = \frac{1}{N_i+1}\left(q_{lm}(i)+\sum_{j=1}^{N_i}q_{lm}(j)\right).
\label{eq:3}
\end{equation}
The $l$-fold local BOO parameter $Q_l$ is then defined by averaging over the $m$-values associated with a given $l$:
\begin{equation}
Q_l(i) = \sqrt{\frac{4\pi}{2l+1}\sum_{m=-l}^l|Q_{lm}(i)|^2}.
\label{eq:4}
\end{equation}

In Fig.~\ref{fig2}, we show the probability distribution of $Q_6$ for metallic glass samples of Cu$_{50}$Zr$_{50}$ prepared using the LJ model at cooling rate $R=2.6 \times 10^{11} \rm K/s$ and using the EAM model at $R=1.0\times10^{10}\rm K/s$, as well as disordered FCC samples for both the LJ and EAM models. The thermally quenched LJ and EAM samples possess similar $P(Q_6)$; both are disordered with a peak near $Q_6 \approx 0.12$.  $P(Q_6)$ for the disordered FCC systems with LJ interactions is broader than those for the thermally quenched samples and the peak is shifted to $Q_6 \approx 0.15$. $P(Q_6)$ for the disordered FCC systems with EAM interactions has a peak near $Q_6 \approx 0.2$, and it possesses a broad tail that extends beyond $Q_6 \sim 0.5$.  Thus, the FCC-initialized systems with EAM interactions include a significant number of crystal-like atoms. 

\begin{figure}[tbp]
\centering
\includegraphics[width=0.48\textwidth]{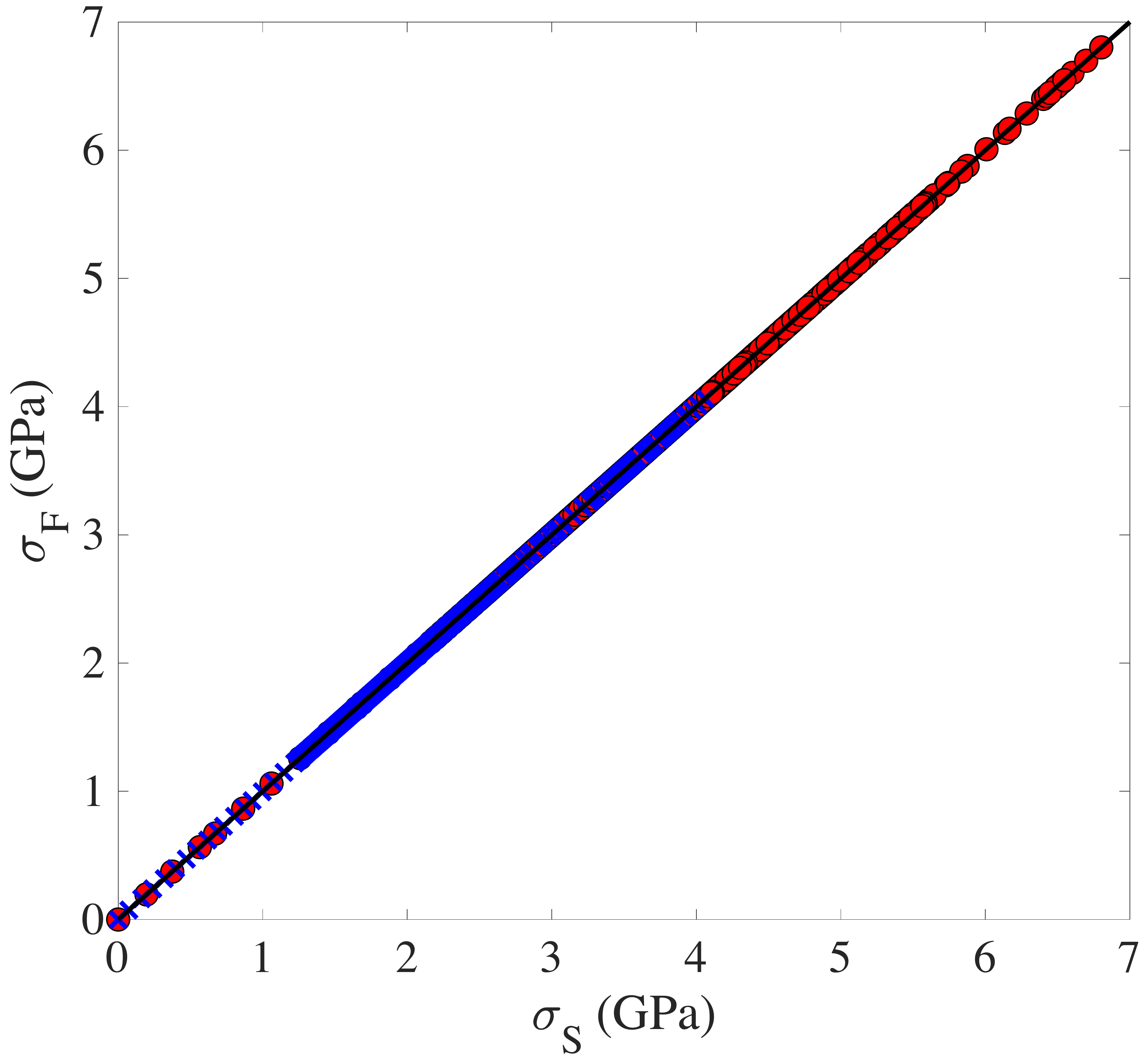}
\caption{Two definitions of the engineering stress $\sigma_F=F_{zz'} L_z/A_0$ and $\sigma_{\cal S}=S_{zz}/A_0$ plotted against each other. $F_{zz'}$ is the total force in the $z$-direction crossing the $z'$ plane (Eq.~(\ref{force})) and $S_{zz}$ is the $zz$-contribution to the virial $S_{zz}$ (Eq.~(\ref{virial})). $A_0$ is the undeformed area of the sample and $L_z$ is its length in the $z$-direction. Data from uniaxial tension tests on Cu$_{50}$Zr$_{50}$ modeled using LJ (red circles) and EAM interactions (blue crosses) are included. The solid black line indicates that $\sigma_F=\sigma_{\cal S}$.}
\label{fig3}
\end{figure}

\begin{figure}
\centering
\includegraphics[width=0.425\textwidth]{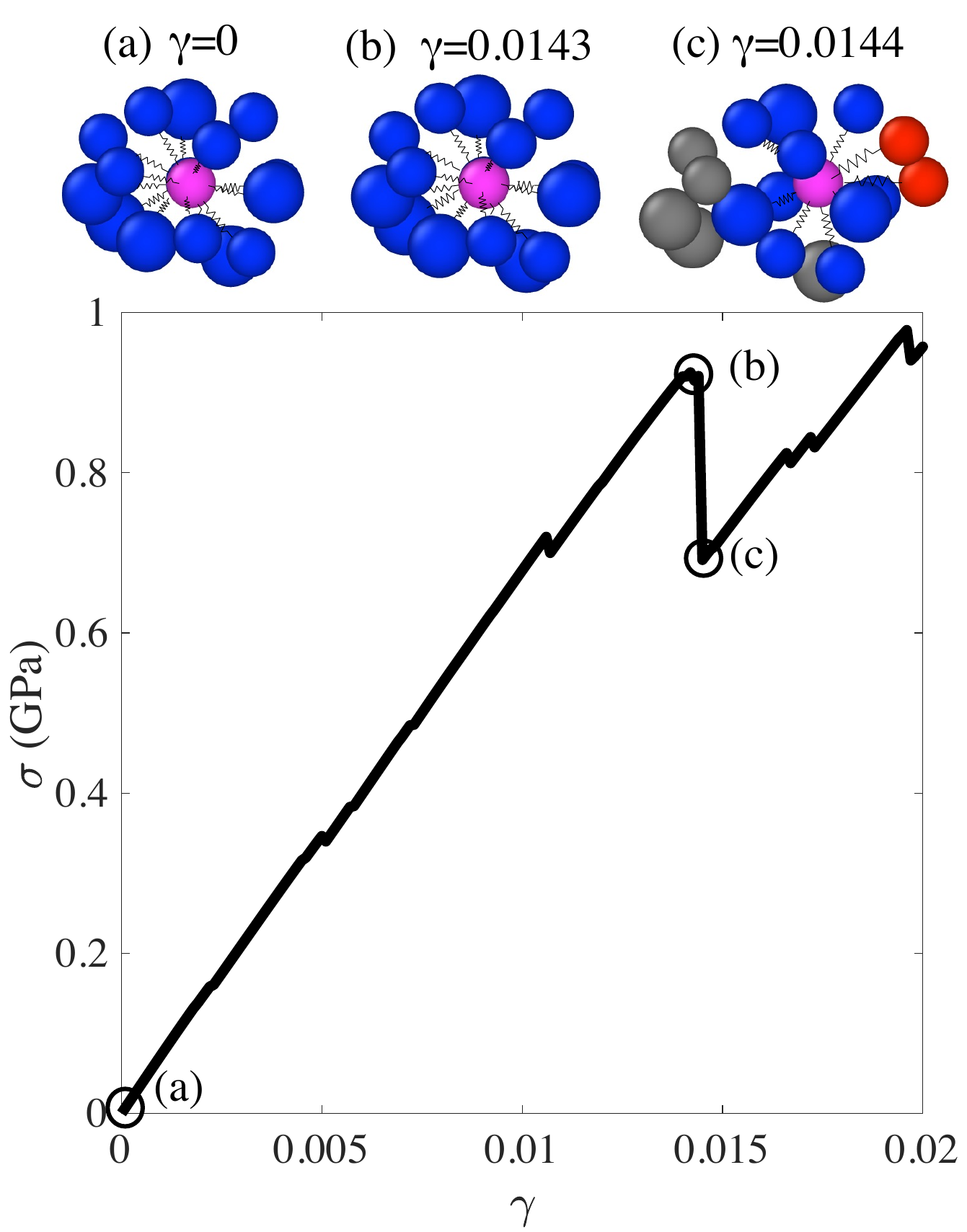}
\caption{Engineering stress $\sigma$ as a function of strain $\gamma$ from a uniaxial tension test of a $\rm Cu_{50}Zr_{50}$ sample generated using the EAM model at cooling rate $R = 10^{10} {\rm K/s}$. The local regions in the three insets (a)-(c), which show the nearest (Voronoi) neighbors of a specified central atom (magenta), are taken from systems with total engineering stress labeled (a)-(c) in the main plot. Between total strain (a) $\gamma=0$ and (b) $0.0143$, there is no change in the (blue) nearest neighbors of the selected central atom. After a large atomic rearrangement event at (c) $\gamma=0.0144$, some of the atoms that were nearest neighbors of the central atom in (a) and (b) are no longer nearest neighbors (grey). The central atom also gains new nearest neighbors (red) that were not nearest neighbors at $\gamma=0$.}
\label{fig4}
\end{figure}

\begin{figure*}
\includegraphics[scale=0.75]{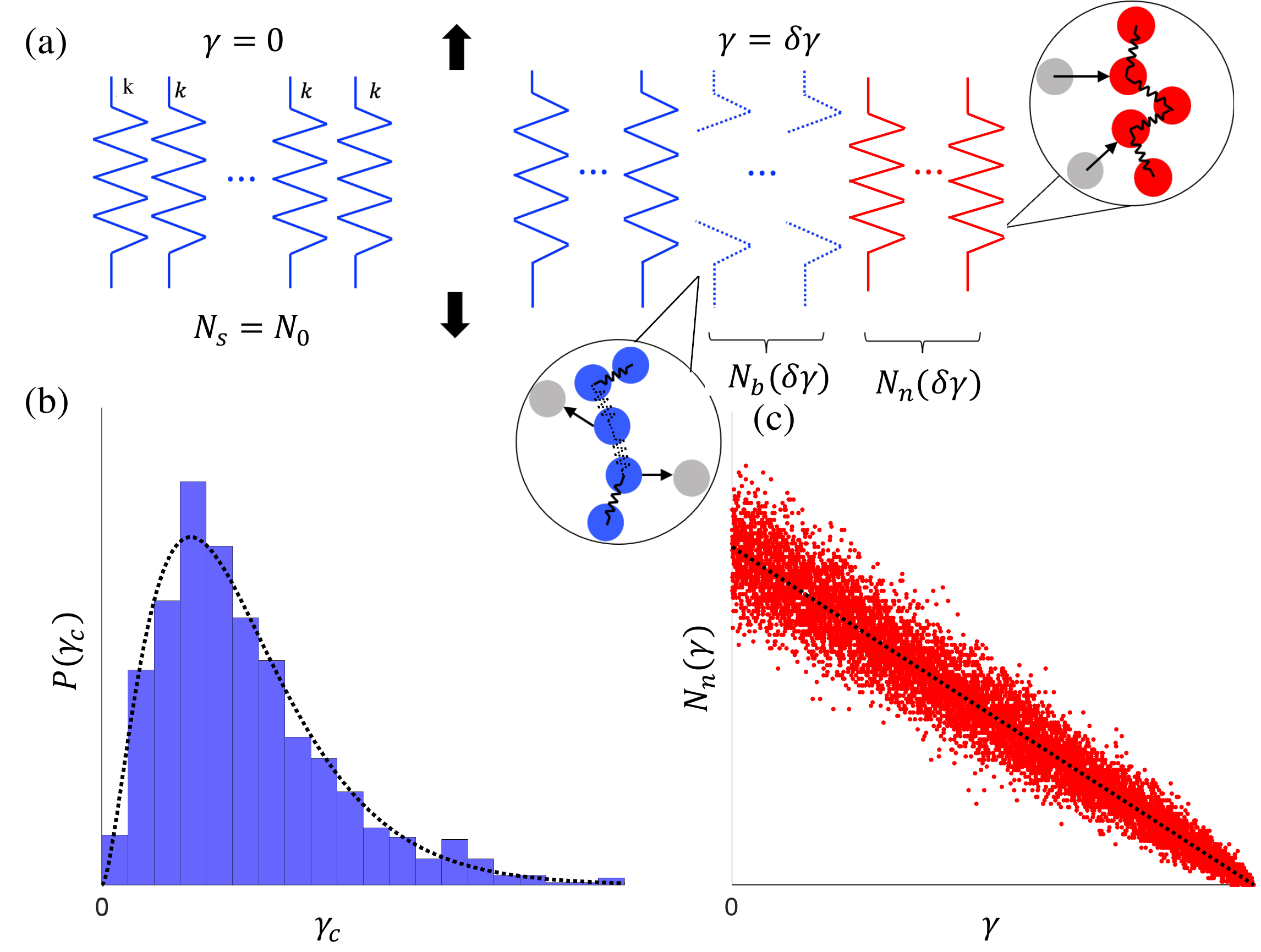}
\caption{(a) Schematic of the spring network model. In the undeformed sample at $\gamma=0$, $N_s=N_{0}$ identical springs are initially connected in parallel with spring constant $k$ and can transmit force in the (vertical) pulling direction. After a single step in uniaxial strain $\gamma=\delta\gamma$, $N_{b}(\delta\gamma)$ springs break according to each spring's threshold strain $\gamma_c$, which is drawn randomly from a Gamma distribution. The breaking of a spring, which represents an atomic rearrangement event (atoms change color from blue to gray), prevents that spring from transmitting force in the pulling direction.  Also, during an atomic rearrangement event, $N_n(\delta\gamma)$ springs can form and begin to transmit force (atoms change color from gray to red). (b) The probability distribution of spring breaking threshold strains $P(\gamma_c)$, which converges to a Gamma distribution in the $N_{0} \rightarrow \infty$ limit (dotted line). In this example, the two shape parameters for $P(\gamma_c)$ are $\alpha=2.7$ and $\beta=0.02$.  (c) The number of new springs $N_{n}(\gamma)$ as a function of strain $\gamma$. At each strain, $N_n(\gamma) \sim N_{n}^{p}(\gamma)p$ springs form on average, where $N_{n}^{p}(\gamma)$ is the number of potential new springs that can form and $p$ is the probability that these new springs are instantiated. We show data for $p=0.1$ and $N_n^p(\delta \gamma)=10^3$. In the $N_{n}^{p} \rightarrow \infty$ limit, $N_{n}(\gamma)$ converges to the black dotted line with vertical intercept $pN_n^p(\delta \gamma)$ and slope $p(dN_{n}^{p}/d\delta\gamma)$.}
\label{fig5}
\end{figure*}

\subsection{Athermal, quasistatic uniaxial tension}

After generating the Cu$_{50}$Zr$_{50}$ metallic glass samples, we perform athermal, quasistatic uniaxial tension tests. In particular, we apply successive small uniaxial strain steps of $\delta\gamma=10^{-4}$ along the $z$-direction by increasing the sample length from its current value $L_z$ to $L_z' = L_{z} + \Delta L_z$ and shifting the $z$-positions of the atoms affinely such that $z'_i = z_i(1+\delta \gamma)$ (as shown in the inset of Fig.~\ref{fig1} (b)). Each strain step is followed by potential energy minimization. $\Delta L_z = L_{z0} \delta\gamma$ and $L_{z0}$ is the original length of the sample in the $z$-direction. Before applying the tensile deformations, we open the boundaries in the $x$- and $y$-directions to allow necking of the sample. To remove the residual stress caused by opening the boundaries, we apply athermal, quasistatic tension or compression in the $z$-direction until the engineering stress is zero. We then apply athermal, quasistatic tensile deformation in small strain steps until the total strain reaches $\gamma=1$.

\begin{figure}[tbp]
\centering
\includegraphics[width=0.48\textwidth]{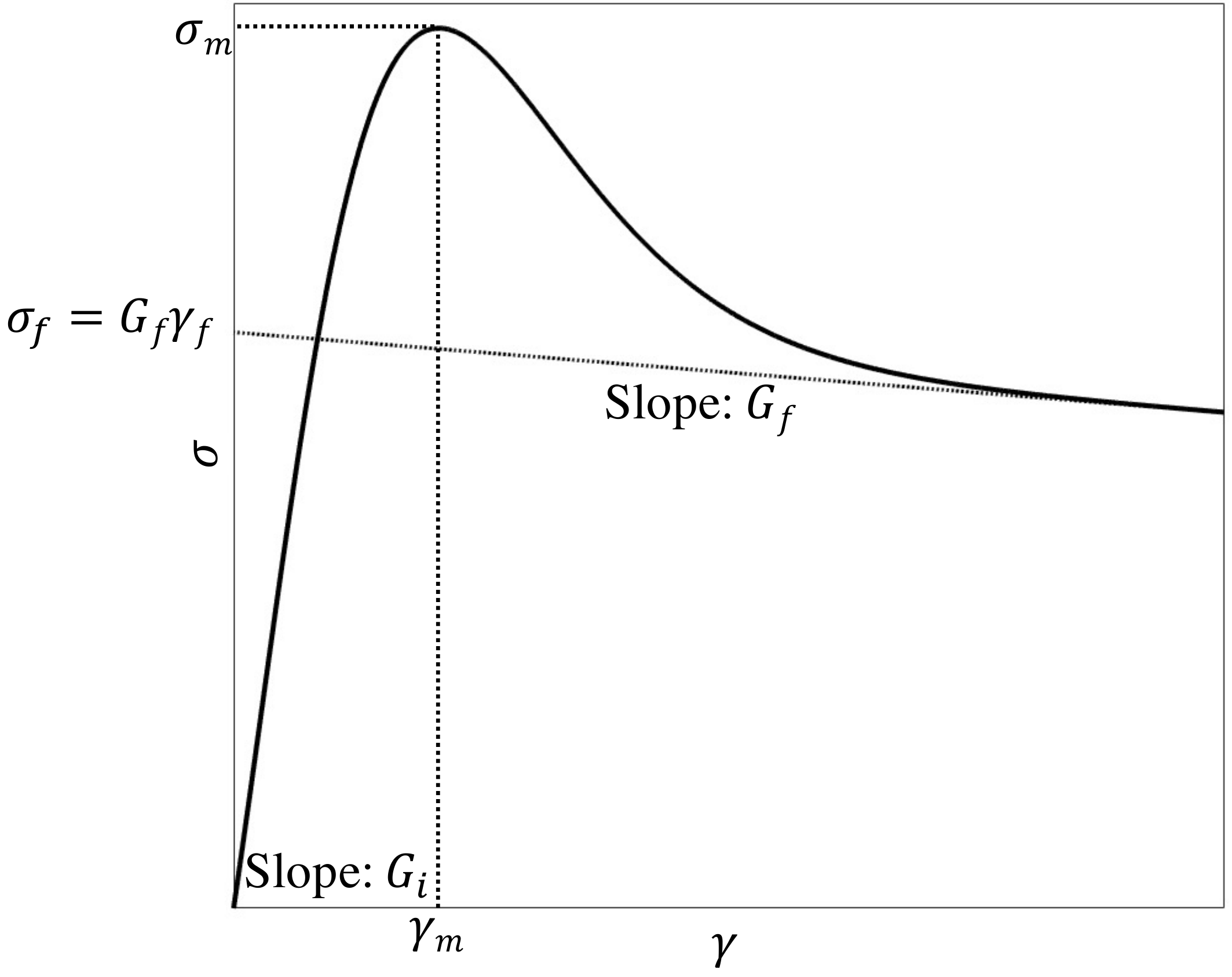}
\caption{Schematic of the engineering stress $\sigma$ versus strain $\gamma$ during uniaxial tension, including the definitions of the five parameters that characterize the shape of $\sigma(\gamma)$. $G_i$ is the slope of $\sigma(\gamma)$ at $\gamma=0$, $\sigma_{m}$ is the maximum engineering stress, $\gamma_{m}$ is the strain at which the maximum engineering stress occurs, $\sigma(\gamma_m)=\sigma_{m}$, $\gamma_f >0$ is the failure strain at which $\sigma=0$, and $G_f$ is the slope of $\sigma(\gamma)$ at $\gamma=\gamma_f$.}
\label{fig6}
\end{figure}

\subsection{Calculating the Engineering Stress}

During uniaxial tensile deformation, the thickness of the metallic glasses becomes length dependent due to necking of the sample. The true stress is defined as the total force in the $z$-direction divided by the cross-sectional area of the sample at each strain. Thus, the true stress is difficult to calculate since we would need to accurately describe the surface of the deformed sample. In contrast, the engineering stress is defined as the total force in the $z$-direction divided by the undeformed cross-sectional area, which is nearly uniform over the length of the sample.  We show typical engineering stress versus strain curves for thermally quenched Cu$_{50}$Zr$_{50}$ metallic glasses modeled using the EAM potential in Fig.~\ref{fig1} (b).

To calculate the engineering stress, we consider the total force in the $z$-direction crossing the $x$-$y$ plane in the sample~\cite{RN14, RN63} at $z=z'$. Note that any $x$-$y$ plane in the sample gives the same total force in the $z$-direction because the system is in force balance. 
The total force in the $z$-direction crossing the $z'$ plane is
\begin{equation}
\label{force}
F_{zz'}= \sum_{i,j} F_{ijz}, 
\end{equation}
where $F_{ijz}$ is $z$-component of the force ${\vec F}_{ij}$ on atom $i$ from other atoms $j$, and the sum only includes atom pairs $i$ and $j$ such that ${\vec r}_{ij}$ intersects the plane $z=z'$. Since ${\vec F}_{ij}=-{\vec F}_{ji}$, the sum of $F_{ijz}$ only includes the force on atom $i$ or $j$ that has the lower $z$-coordinate than the other atom. We define our coordinate system such that $-L_z/2 \le z \le L_z/2$, and thus we set $z'=0$. Since $L_z/2 > r_c$ for both the LJ and EAM models, atoms only interact across the plane $z'=0$ and not through the periodic image cells. The total force in the $z$-direction across the plane $z'=0$ is the $z$-component of the total force on atoms in the lower region ($z' < 0$) arising from interactions with atoms in the upper region ($z' >0$). 

For LJ interactions, ${\vec F}_{ij} = (dU/dr_{ij}) {\hat r}_{ij}$. For EAM interactions, the force on atom $i$ from $j$ is
\begin{eqnarray}
\label{force_EAM}
{\vec F}_{ij} & = & [(\partial {\cal F}_i/\partial \rho_{i}) (\partial \rho_{i}/\partial r_{ij})+(\partial {\cal F}_j/\partial \rho_{j})(\partial \rho_{j}/\partial r_{ij}) \nonumber \\
& & +(\partial \phi^p_{ij}/\partial r_{ij})]{\hat r_{ij}},
\end{eqnarray}
which includes two additional terms arising from the embedding function ${\cal F}_i$ and where $\rho_i=\sum_{j \ne i} \rho_{ij}(r_{ij})$ is the electron density of atom $i$ at position ${\vec r}_i$ and $\phi^p_{ij}$ is the EAM pair potential energy function.  We then define the engineering stress, $\sigma_F = F_{zz'}/A_0$, where $A_0$ is the cross-sectional area of the undeformed sample. We first determine the $\alpha$-shape of the undeformed sample, and then calculate its volume $V_0$ and $A_0 = V_0/L_{z0}$. We find that the relative fluctuations in $A_0$ are less than $0.6\%$ for all preparation protocols and both LJ and EAM interactions. 

We also compared the results for $\sigma_F$ to the results for the engineering stress obtained using the $zz$-component of the virial stress tensor:
\begin{equation}
\label{virial}
S_{zz} = \sum_{i=1}^N z_i F_{iz}.
\end{equation}
In terms of $S_{zz}$, the engineering stress is defined as 
\begin{equation}
\label{virial_force}
\sigma_{\cal S}= \frac{S_{zz}}{L_{z0}A_0}. 
\end{equation}
In Fig.~\ref{fig3}, we show that, as expected, $\sigma_F =\sigma_{\cal S}$ for uniaxial tension applied to Cu$_{50}$Zr$_{50}$ metallic glass samples modeled using the LJ and EAM interactions. Below, we use $\sigma \equiv \sigma_F$ to display the engineering stress.

\subsection{Spring network model}

We now describe the key elements of the spring network model, which are illustrated in Fig.~\ref{fig4}.  In the undeformed state at $\gamma=0$, the sample is in mechanical equilibrium, the atoms are in their equilibrium positions with an initial set of nearest (Voronoi) neighbors, and the engineering stress $\sigma$ is zero.  As the uniaxial strain $\gamma$ is applied, $\sigma$ increases roughly linearly. Small drops in $\sigma$ occur prior to point (b), but the nearest neighbors of the selected central atom remain the same as they were at $\gamma=0$. Between points (b) and (c), a large engineering stress drop occurs, which corresponds to an atomic rearrangement event where five atoms are no longer nearest neighbors of the central atom, and two new atoms become nearest neighbors of the central atom. Changes in the nearest neighbors of atoms can occur throughout the sample and at various strain steps during uniaxial tension deformation. Inspired by these atomic rearrangement events, we develop a mesoscopic one-dimensional spring network model that considers the breaking and forming of springs in parallel during deformation. The breaking (formation) of a spring represents a loss (gain) of nearest neighbor atoms in a local region. 
The total force (in the $z$-direction) resisting the extension of the sample will be related to the number of springs at each strain step and how much each spring is stretched. Our goal is to develop a simple model with a small number of parameters, yet we want it to possess a sufficient number of parameters so that it is able to quantitatively characterize the mechanical response of metallic glasses. 

\begin{figure*}
\centering
\includegraphics[scale=0.58]{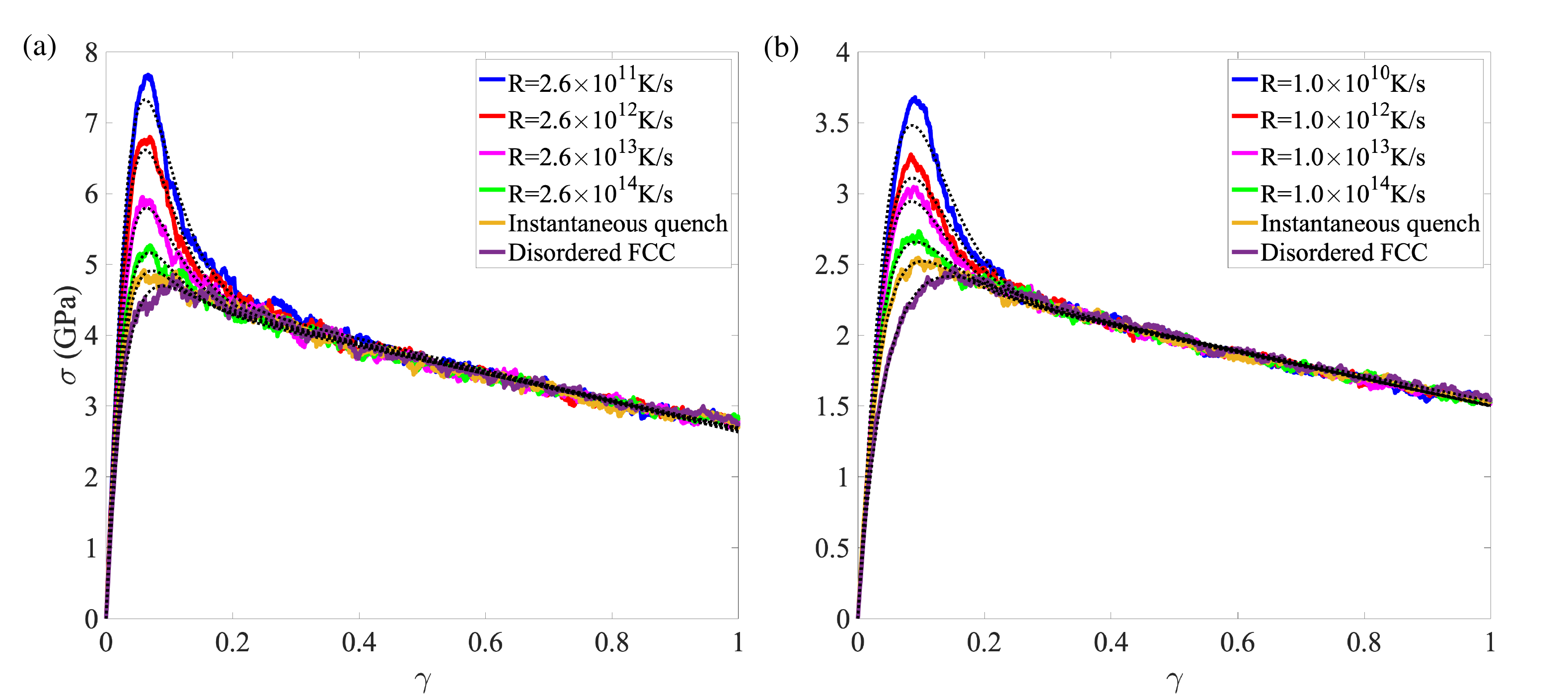}
\caption{Engineering stress $\sigma$ plotted as a function of strain $\gamma$ during uniaxial tension deformation of Cu$_{50}$Zr$_{50}$ metallic glass samples (solid lines) obtained using the (a) LJ and (b) EAM models for several cooling rates, the instantaneous thermal quenches, and the disordered FCC structures averged over $50$ samples. The dotted lines represent best fits of $\sigma(\gamma)$ to the prediction from the spring network model (Eq.~\ref{predicted_stress}) using the parameters in Fig.~\ref{fig9}.}
\label{fig7}
\end{figure*}

The spring network model is summarized in Fig.~\ref{fig5} (a). Initially, we assume that there are $N_s=N_0$ identical, unstretched springs in parallel, each with spring constant $k$ and rest length $l_0$. After each applied step strain, we assume that all springs experience the same amount of strain, $\delta\gamma$. Thus, the force experienced by spring $j$ after a total strain $\gamma^j$ is $F^j = kl_0 \gamma^j$. Each spring is assigned a cutoff strain $\gamma_{c} >0$, which is the total strain at which the spring breaks. $\gamma_c$ is randomly selected from a Gamma distribution:
\begin{equation}
P(\gamma_{c}) = \frac{\beta^\alpha}{\Gamma(\alpha)}\gamma_{c}^{\alpha-1}e^{-\beta\gamma_c},
\label{eq:8}
\end{equation}
where $\Gamma(.)$ is the Gamma function. $P(\gamma_c)$ has two parameters $\alpha$ and $\beta$ that characterize its shape: the mean is $\alpha/\beta$ and the variance is $\alpha/\beta^2$. (See Fig.~\ref{fig5} (b).)  Selecting cutoff strains from $P(\gamma_c)$ gives a coarse-grained representation of the structural disorder in metallic glasses, such as fluctuations in the values of ${\vec r}_{ij} \cdot {\hat z}$ for interacting atomic pairs. When a given spring breaks, the force on that spring is set to zero. 

At each strain step, new springs can also form. We assume that the number of potential new springs decreases linearly with the total strain,
\begin{equation}
N_{n}^{p}(\gamma_{n})=N_{n}^{p}(\delta \gamma)-\frac{dN_{n}^{p}}{d\delta\gamma}\gamma_{n-1}, 
\label{eq:6}
\end{equation}
where $\gamma_n=n\delta \gamma$ is the total strain after $n$ strain steps. This form for $N_{n}^{p}(\gamma_{n})$ is consistent with the fact that the cross-sectional area decreases with increasing uniaxial strain, and thus there are fewer atoms in the transverse direction for forming new spring connections. (We also considered quadratic and piecewise linear functions for $N^p_{n}(\gamma_n)$, which include an additional parameter. However, the form in Eq.~\ref{eq:6} provided high-quality fits to the data from simulations of athermal, quasistatic uniaxial tension.) The {\it potential} new springs form instantiated new springs with probability $p$. The scatter plot in Fig.~\ref{fig5} (c) shows the number of new springs $N_n(\gamma_n)=p N_{n}^{p}(\gamma_n)$ as a function of total strain $\gamma_n$ for $p=0.1$ and $N_n^p(\delta \gamma)=10^3$. When a new spring is formed, it is initiated with the same rest length $l_0$ and $\gamma_{c}$ is randomly selected from the same $P(\gamma_c)$ regardless of the total strain. Newly formed springs also experience the same incremental strain $\delta \gamma$ at each applied strain step. 

The total force of the spring network at total strain $\gamma_n=n\delta \gamma$ is 
\begin{equation}
F_{s}(\gamma_n)= \sum_{j=1}^{N_{s}}kl_0\gamma_n^j,
\label{eq:7}
\end{equation}
where $\gamma_n^j$ is the total strain experienced by the $j$th spring after $n$ strain steps applied to the initial undeformed sample, which differs from the total strain $\gamma_n$ for springs that were not present at $\gamma=0$.  Thus, to calculate the total force for the spring network model at $\gamma_n$, we need to track the number of initial springs that are still intact at $\gamma_n$, the number of new springs that have formed since $\gamma=0$, and at what strains each of these new springs form. 

The engineering stress versus strain curve $\sigma(\gamma)$ for uniaxial tension for a single Cu$_{50}$Zr$_{50}$ metallic glass sample using the EAM model with $N=3456$ atoms includes numerous rapid drops in stress as shown in Fig.~\ref{fig4}.  Several studies have shown that for metallic glasses undergoing athermal, quasistatic deformation, the size of the stress drops decreases and the number of stress drops increases with increasing system size. To mimic the large-system limit, we can calculate the ensemble-averaged $\sigma(\gamma)$ from the numerical simulations of the LJ and EAM models over many realizations.  The results for $\sigma(\gamma)$ from the spring network model in the large-system limit can be obtained by taking the limits $N_s \rightarrow \infty$ and $N_n^p \rightarrow \infty$. In this limit, we can derive an analytical expression for the total force in the spring network model (Eq.~(\ref{eq:7})), and compare it to $\sigma(\gamma) A_0$ obtained from the athermal, quasistatic tension tests of Cu$_{50}$Zr$_{50}$ samples modeled using the LJ and EAM interaction potentials. 

If $N_0$ springs are initialized at $\gamma=0$, the number of these springs that have not broken after strain $\gamma_n$ is
$N(\gamma_n)= N_0[1-C(\gamma_n)]$, where $C(\gamma_n)$ is the cumulative distribution function for the Gamma distribution. Since these springs experience the same total strain, we can calculate the total force from the remaining springs at $\gamma_n$:
\begin{equation}
F_{s0}(\gamma_n) = kl_0\gamma_n N_0[1-C(\gamma_n)]. 
\label{eq:9}
\end{equation}

In the limit $N_{n}^{p} \rightarrow \infty$, the number of new springs that form, $N_{n}(\gamma_n)$, can also be derived. Using Eq.~(\ref{eq:6}), we find
\begin{equation}
\begin{split}
N_{n}(\gamma_n)&=p N^p_n(\gamma_n)\\
&=N_{n}(\delta \gamma)-\frac{dN_{n}}{d\delta\gamma}\gamma_{n-1}. 
\end{split}
\label{eq:10}
\end{equation}
To calculate the total force from the newly formed springs, we need to take into account the fact that $N_{n}(\gamma_n)$ new springs are generated at $\gamma_n$ and these springs break after an additional strain of $\gamma_c$, which is selected randomly from a Gamma distribution. The total force arising from the new springs at $\gamma_n$ is
\begin{multline}
    F_{sn}(\gamma_n)= kl_0\sum_{i=0}^{n-1} \left(N_{n}(\delta \gamma)-\frac{dN_{n}}{d\delta\gamma}(\gamma_{n-1}-\gamma_i)\right)\\
\times \gamma_i[1-C(\gamma_i)],
\label{eq:11}
\end{multline}
where $F_{sn}(\gamma_0)=0$. Thus, the total force in the spring network at strain $\gamma_n$ is given by 
\begin{equation}
\label{eq12}
F_s(\gamma_n) = F_{s0}(\gamma_n) + F_{sn}(\gamma_n),
\end{equation}
where $F_{s0}$ and $F_{sn}$ are provided in Eqs.~(\ref{eq:9}) and~(\ref{eq:11}), respectively. Taking the continuum limit in strain and normalizing the spring force by the sample's undeformed cross-sectional area $A_0$ allow a comparison to the engineering stress versus strain obtained in the simulations of athermal, quasistatic uniaxiial tensile deformations:
\begin{equation}
\label{predicted_stress}
\sigma_s(\gamma) =F_s(\gamma)/A_0.
\end{equation}

\begin{figure*}
\includegraphics[scale=0.58]{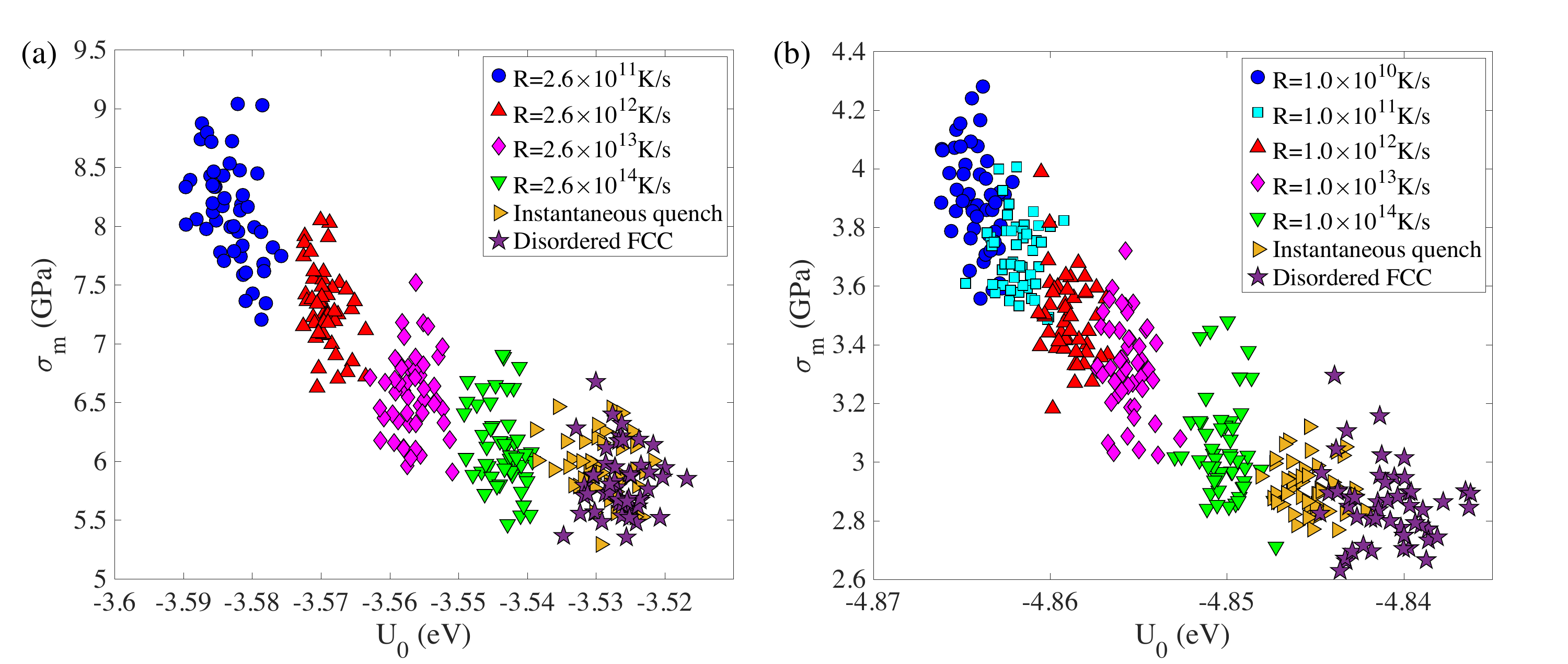}
\caption{Maximum engineering stress $\sigma_{m}$ during athermal, quasistatic tension tests plotted versus the total potential energy per atom $U_0$ at $\gamma=0$ for $50$ Cu$_{50}$Zr$_{50}$ metallic glass samples modeled using the (a) LJ and (b) EAM potentials. The rightward triangles and stars represent instantaneous thermal quenches and disordered FCC structures and the other symbols represent different cooling rates $R$ used to prepare the samples.} 
\label{fig8}
\end{figure*}

\subsection{Comparison of the Predictions of the Spring Network Model and Results from Atomistic Simulations}
\label{comparison}

The engineering stress $\sigma_s$ in Eq.~(\ref{predicted_stress}) from the spring network model has five parameters. These include the number of initial springs $N_{0}$, number of new springs that are formed during the first strain step $N_{n}(\delta\gamma)$, and the change in the number of new springs that are formed per strain step $dN_{n}/d\delta\gamma$. In addition, the cumulative distribution $C(.)$ that controls the cutoff strain is characterized by two shape parameters, $\alpha$ and $\beta$.  Motivated by $\sigma(\gamma)$ from experiments and simulations of uniaxial tension of metallic glasses in Fig.~\ref{fig1}, we can now relate these five parameters from the spring network model to five key features of the shape of the engineering stress versus strain curve illustrated in Fig.~\ref{fig6}: 1) the slope of the engineering stress $G_i=d\sigma_s/d\gamma$ at $\gamma=0$, 2) the ultimate strength $\sigma_m=\sigma_s(\gamma_m)$, 3) the strain $\gamma_m$ at ultimate strength, 4) the failure strain $\gamma_f$ at which $\sigma_s(\gamma_f)=0$, and 5) the slope of the engineering stress $G_f = d\sigma_s/d\gamma$ at $\gamma_f$.  The following five equations,
\begin{subequations}
  \begin{equation}
  \frac{d \sigma_{s}(0)}{d \gamma} -G_i = 0 
  \end{equation}
  \begin{equation}
 \sigma_s(\gamma_{m}) - \sigma_{m} = 0 
 \end{equation}
 \begin{equation}
 \frac{d \sigma_{s}(\gamma_{m})}{d\gamma} = 0
 \end{equation}
 \begin{equation}
 \sigma_{s}(\gamma_{f})= 0
 \end{equation}
 \begin{equation}
 \frac{d \sigma_{s}(\gamma_{f})}{d \gamma} -G_f = 0 
 \end{equation}
 \label{eq:19}
\end{subequations}

\noindent can be used to express the parameters in the spring network model, $N_{0}$, $N_{n}(\delta \gamma)$, $dN_{n}/d\delta\gamma$, $\alpha$, and $\beta$, in terms of the shape features of $\sigma_s(\gamma)$, i.e., $G_i$, $\sigma_m$, $\gamma_m$, $\gamma_f$, and $G_f$.  We then use a Levenberg-Marquardt nonlinear least squares algorithm to find the optimal values of $G_i$, $\sigma_m$, $\gamma_m$, $\gamma_f$, and $G_f$ such that $\sigma_s(\gamma)$ matches $\sigma(\gamma)$ from the athermal, quasistatic tension simulations, as well as from uniaxial tension tests of Zr-based metallic glasses in experiments. 

\begin{figure*}
\centering
\includegraphics[scale=0.68]{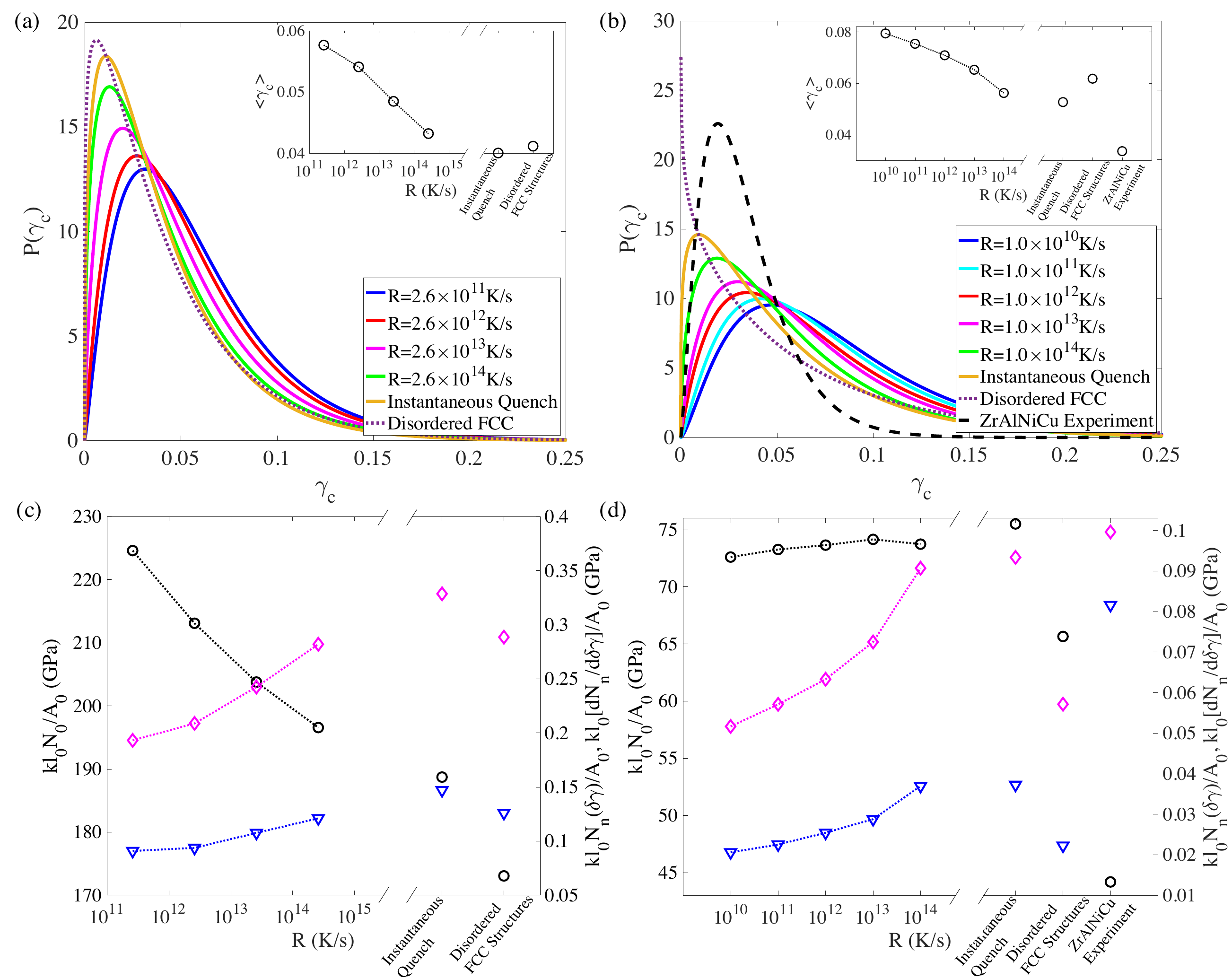}
\caption{The optimal values of the parameters of the spring network model obtained from best fits of $\sigma_s(\gamma)$ to $\sigma(\gamma)$ from the athermal, quasistatic uniaxial tension simulations of the LJ [(a) and (c)] and EAM [(b) and (d)] models of Cu$_{50}$Zr$_{50}$. Panels (b) and (d) also show optimal values of the parameters for $\sigma(\gamma)$ obtained from experimental studies of uniaxial tension on $\rm Zr_{65}Al_{10}Ni_{10}Cu_{15}$ metallic glasses performed at $\rm 593K$~\cite{RN111}. $\sigma(\gamma)$ from these experiments is shown in Fig.{~\ref{fig12}}(a). Panels (a) and (b) and their insets show the variation of $P(\gamma_c)$ and the average $\langle \gamma_c \rangle$ with the sample preparation protocols and experimental studies. In the main panels (a) and (b), the thermal quenches, disordered FCC samples, and experimental studies are represented by solid, dotted, and dashed lines, respectively. Panels (c) and (d) show $kl_0 N_0/A_0$ (circles) on the left vertical axis, and $kl_0 N_n(0)/A_0$ (diamonds) and $k l_0 (dN_n/d\delta\gamma)/A_0$ (triangles) on the right vertical axis for the different sample preparation protocols and experimental studies.} 
\label{fig9}
\end{figure*}

\section{Results}
\label{results}

In this section, we first describe the results from athermal, quasistatic simulations of the mechanical response of the LJ and EAM models of Cu$_{50}$Zr$_{50}$ metallic glasses to uniaxial tension. We also show the correlation between the maximum engineering stress during uniaxial tension and the potential energy per atom of the undeformed samples. We then compare the simulation results for $\sigma(\gamma)$ from athermal, quasistatic uniaxial tension to the prediction of the engineering stress versus strain from the spring network model. We show the dependence of the five parameters of the spring network model on the sample preparation protocol and relate these parameters to key features of the shape of $\sigma(\gamma)$. Lastly, we show best fits of the spring network model to the results of experiments on uniaxial tension applied to several Zr-based metallic glasses, including ${\rm Zr}_{65}{\rm Al}_{10}{\rm Ni}_{10}{\rm Cu}_{15}$ , $\rm Cu_{49}Zr_{51}$, and $\rm Zr_{56}Ni_{22}Al_{22}$. 

\subsection{Engineering Stress versus Strain from Athermal, Quasistatic Uniaxial Tension}

In Fig.~\ref{fig7}, we show $\sigma(\gamma)$ for the LJ and EAM models of Cu$_{50}$Zr$_{50}$ metallic glass samples obtained from thermal quenches over a range of cooling rates, instantaneous thermal quenches, and disordered FCC structures. The LJ and EAM models show qualitatively similar mechanical response over the full range of strain. For all systems, at small strains, the engineering stress increases approximately linearly with strain. The slope $d\sigma/d\gamma=G_i$ at $\gamma=0$ is only weakly dependent on the preparation protocol of the metallic glasses. At larger strains, $\sigma(\gamma)$ becomes nonlinear and reaches a peak engineering stress $\sigma_m$ that grows monotonically with decreasing cooling rate, i.e., $\sigma_m$ is smallest for samples generated via instantaneous thermal quenches and is the largest for samples generated via the slowest cooling rates. 

The disordered FCC structures possess the smallest $\sigma_m$ of the systems in Fig.~\ref{fig7} and the strain at which the peak stress occurs is shifted to larger strains $\gamma_m \sim 0.2$ compared to the peak strains for the thermally quenched systems. At first, these results may seem counterintuitive.  For example, it is well-known that many crystalline structures possess large peak stress at small strains and can be brittle with a rapid decrease in stress near failure.  However, as shown in Fig.~\ref{fig2}, the disordered FCC structures have the largest $Q_6$ values, yet they are the most ductile of the systems we considered. 

To illustrate why disordered FCC structures have a ductile-like response, we show a scatter plot of $\sigma_m$ during uniaxial tension versus the total potential energy per atom $U_0$ of the corresponding undeformed samples in Fig.~\ref{fig8}. For both the LJ and EAM models, $\sigma_m$ decreases with $U_0$ on average.  In particular, we find that the disordered FCC structures possess the smallest $U_0$ and $\sigma_m$ values. The results for the thermally quenched systems are consistent with prior results on metallic glasses. In particular, numerous studies have found that the fictive temperature and average potential energy per atom decrease with decreasing cooling rate~\cite{RN89} and that increasing the fictive temperature increases the ductility of metallic glasses~\cite{RN56,RN87,RN67}. In the current studies, we have shown that the disordered FCC structures possess higher fictive temperatures (or $U_0$) than the samples that were prepared via instantaneous thermal quenches. Another important feature of $\sigma(\gamma)$ for both the LJ and EAM models of Cu$_{50}$Zr$_{50}$ is that it decreases roughly linearly with strain at large strains with slope $G_f < G_i$. Further. $\sigma(\gamma)$ is independent of the preparation protocol at large strains. 

\begin{figure}[tbp]
\centering
\includegraphics[width=8cm]{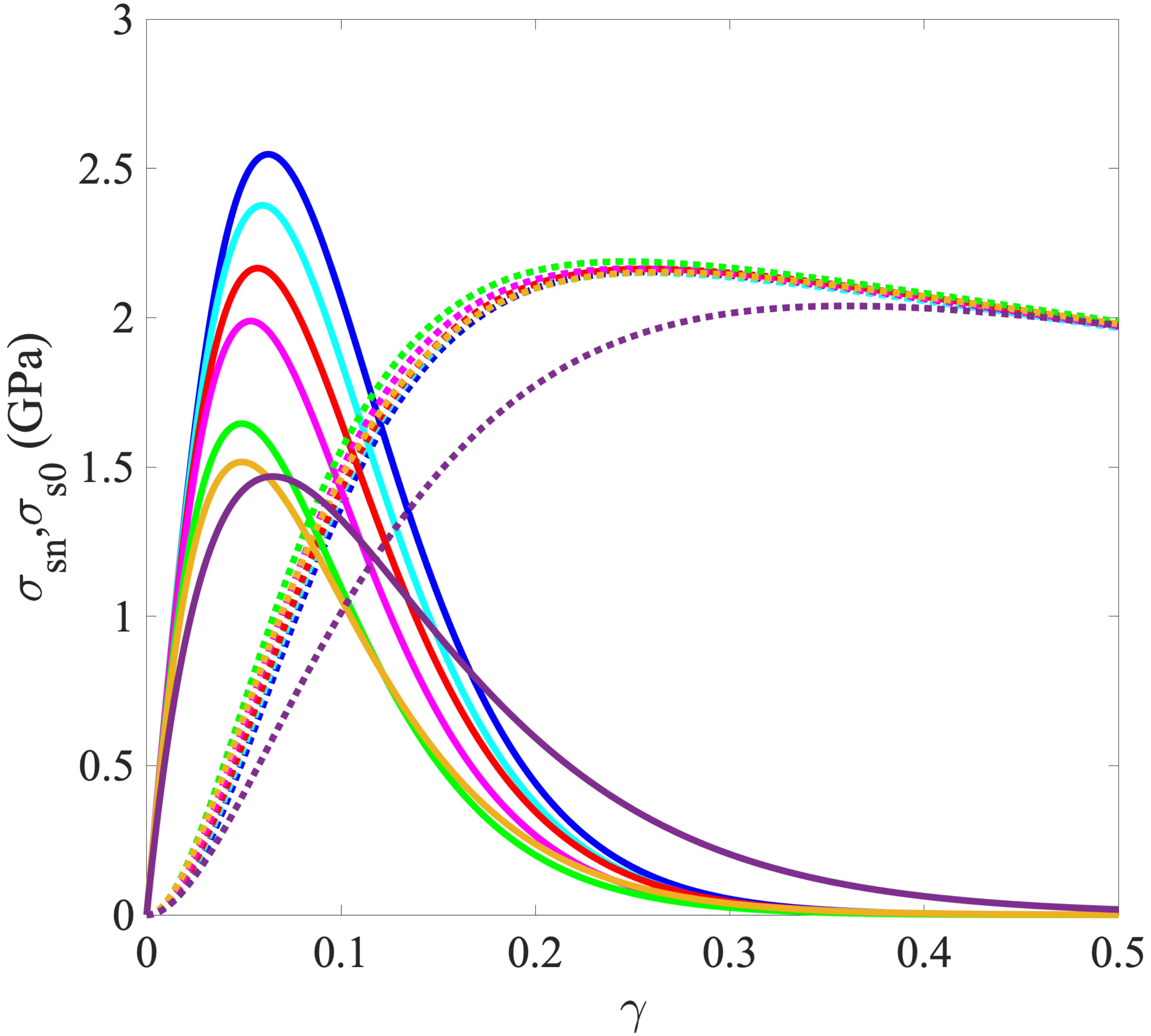}
\caption{$\sigma_s(\gamma)$ from the spring network model (Eq.~\ref{eq12}) can be decomposed into contributions from the initial springs $\sigma_{s0}$ (solid lines) and new springs $\sigma_{sn}$ (dotted lines). We show results from best fits to the EAM model for Cu$_{50}$Zr$_{50}$ samples generated using cooling rates $R=10^{10} {\rm K/s}$ (blue), $10^{11} {\rm K/s}$ (cyan), $10^{12} {\rm K/s}$ (red), $10^{13} {\rm K/s}$ (magenta), and $10^{14} {\rm K/s}$ (green), instantaneous quenches (yellow), and disordered FCC structures (purple).  }
\label{fig10}
\end{figure}

\begin{figure*}
\centering
\includegraphics[scale=0.6]{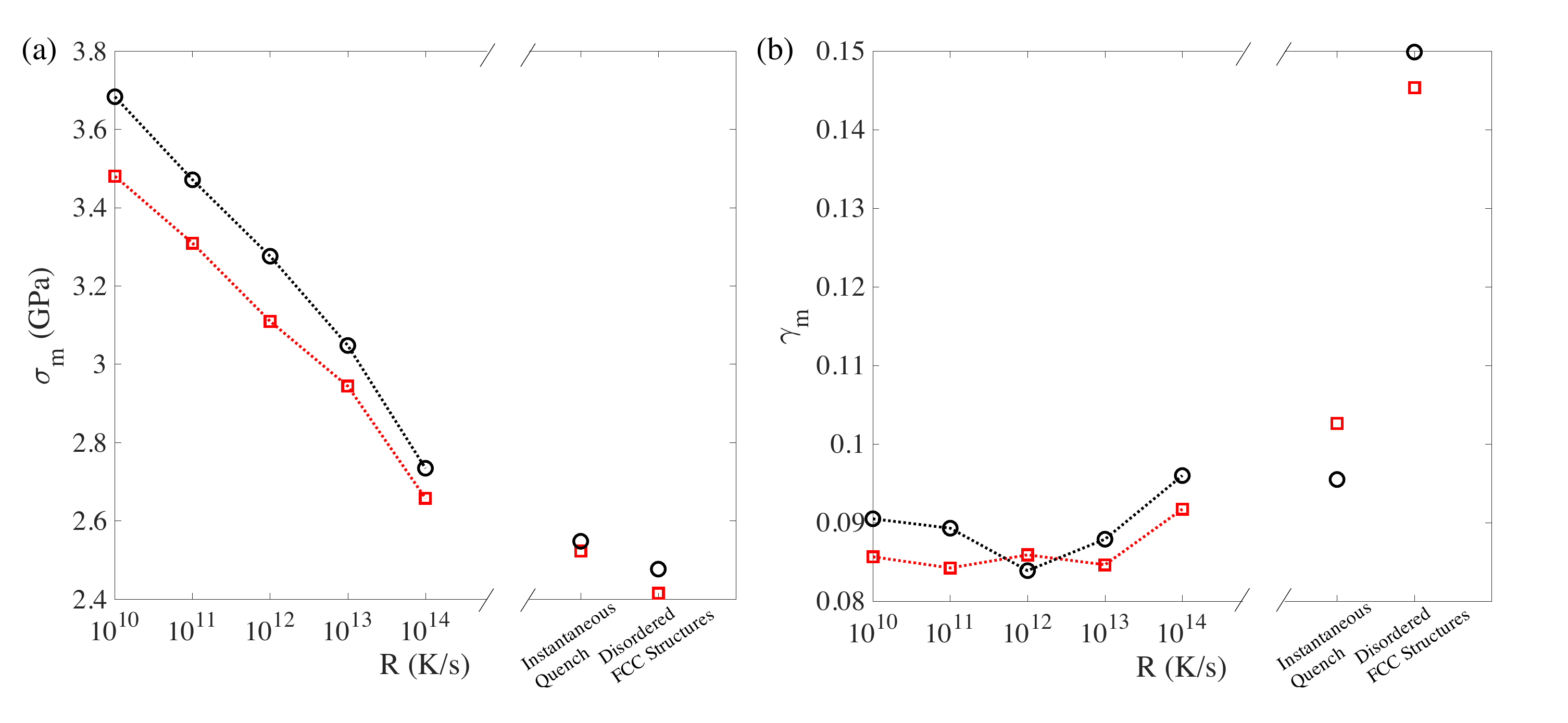}
\caption{(a) Maximum engineering stress $\sigma_{m}$ and (b) the strain $\gamma_{m}$ at which $\sigma(\gamma_m)=\sigma_m$ calculated directly from the spring network parameters (Eq.~\ref{eq:19}) obtained from best fits of the data from the simulations of athermal, quasistatic uniaxial tension of Cu$_{50}$Zr$_{50}$ modeled using the EAM model (red squares). The black circles indicate results directly from $\sigma(\gamma)$ obtained from simulations of athermal, quasistatic uniaxial tensile deformations.}
\label{fig11}
\end{figure*}

\subsection{Comparisons to Prediction of Spring Netowrk Model}

In Fig.~\ref{fig7}, we show the best fits of $\sigma_s$ for the spring network model to $\sigma(\gamma)$ from the simulations of athermal, quasistatic uniaxial tension for the LJ and EAM models of Cu$_{50}$Zr$_{50}$ for all sample preparation protocols.  In general, the fits are high quality for both the LJ and EAM models for all preparation protocols. The only discrepancy occurs for the slowest cooled samples, where the fit slightly underestimates $\sigma_m$. (This small discrepancy can be removed by generalizing the distribution of strain cutoffs $P(\gamma_c)$ to include an additional shape parameter.) 
To assess the quality of the fits of the predicted $\sigma_s(\gamma)$ for the spring network model to $\sigma(\gamma)$ from the atomistic simulations, we calculate the root-mean-square error,
\begin{equation}
\langle \Delta \sigma \rangle= \sqrt{\frac{\sum^n_{i=1}(\sigma(\gamma_i)-\sigma_s(\gamma_i))^2}n},
\end{equation}
where $\gamma_i=i\delta \gamma$ and $n=1/\delta \gamma$. We find that 
the normalized root-mean-square error $\langle \Delta \sigma \rangle/\langle \sigma\rangle \lesssim 0.03 $ for all of the 
simulations of athermal, quasistatic uniaxial tension that we performed.  

The spring network parameters obtained from the fits are displayed in Fig.~\ref{fig9}. In panels (a) and (b), we show how the cutoff distribution $P(\gamma_c)$ and its average $\langle \gamma_c \rangle$ vary with the sample preparation protocol for the LJ and EAM models of Cu$_{50}$Zr$_{50}$. The results are similar for the LJ and EAM models. $P(\gamma_c)$ shifts to larger strains and becomes broader as the cooling rate is {\it decreased}.  The shift and broadening of the distribution toward larger strains indicate that the springs can withstand more elongation before breaking when the sample is prepared at lower cooling rates. These results are consistent with previous simulation studies~\cite{RN55} of shear stress versus strain during athermal, quasistatic simple shear of model metallic glasses. They showed that the frequency of atomic rearrangements and energy loss per rearrangement are reduced for slowly cooled glasses at small strains.  For the disordered FCC structures, $P(\gamma_c)$ is shifted to low values of strain, which indicates that the springs begin to break immediately after the application of uniaxial tension in these samples. 

In Fig.~\ref{fig9} (c) and (d), we show $k l_0 N_0/A_0$, $kl_0 N_{n}(0)/A_0$, and $k l_0(dN_n/d\delta\gamma)/A_0$ for metallic glass samples with different preparation protocols. $k l_0 N_0/A_0$ controls the modulus at $\gamma=0$. For the LJ model, $k l_0 N_0/A_0$ decreases with increasing cooling rate, whereas, $k l_0 N_0/A_0$ is weakly dependent on cooling rate for the EAM model, with a slight decrease for the FCC disordered structures. For both LJ and EAM models of Cu$_{50}$Zr$_{50}$, more rapidly cooled glasses possess larger values of $kl_0 N_{n}(0)/A$ and $k l_0(dN_n/d\delta\gamma)/A$. Thus, more new springs are generated for more rapidly cooled glasses. However, since the cutoff distribution $P(\gamma_c)$ is shifted to smaller strains for more rapidly cooled glasses, these new springs break more frequently. LJ disordered FCC structures possess values of $kl_0 N_{n}(0)/A_0$ and $k l_0(dN_n/d\delta\gamma)/A_0$ that are comparable to the most rapidly cooled metallic glasses, which is consistent with the fact that they have small $\sigma_m$. Instead, $kl_0 N_{n}(0)/A_0$ and $k l_0(dN_n/d\delta\gamma)/A_0$ for the EAM disordered FCC structures are comparable to the values for slowly cooled glasses. 

The predicted stress from the spring network model, $\sigma_s(\gamma)=\sigma_{s0}(\gamma)+\sigma_{sn}(\gamma)$, can be decomposed into contributions from the initial springs $\sigma_{s0}$ that were present at $\gamma=0$ and from the new springs $\sigma_{sn}$ that continue forming after the initial springs break. (See  Fig.~\ref{fig10}.)  We find that the stress contribution from the initial springs, which is controlled by the initial number of springs and the cutoff distribution $P(\gamma_c)$, is strongly protocol dependent. Further, the contribution from the initial springs is large at small strains and decays to zero at large strains. In contrast, $\sigma_{sn}$ is zero at small strains and $\sigma_{sn} \gg \sigma_{s0}$ at large strains. For all thermally quenched samples and for the disordered FCC structures at sufficiently large strains, the stress contribution from the new springs $\sigma_{sn} \sim \sigma_s$ is independent of the sample preparation protocol. This result implies that the variation of $P(\gamma_c)$ with the preparation protocol ({\it cf.} Fig.~\ref{fig9} (a) and (b)) exactly offsets the variation of $kl_0 N_{n}(0)/A_0$ and $k l_0(dN_n/d\delta\gamma)/A_0$ with the preparation protocol ({\it cf.} Fig.~\ref{fig9} (c) and (d)). 

As described in Sec.~\ref{comparison}, we can determine key features of the engineering stress versus strain curves from the parameters in the spring network model.  As an example, in Fig.~\ref{fig11}, we show
the peak stress $\sigma_m$ and the strain $\gamma_m$ at which it occurs determined directly from the five optimal spring network parameters obtained from fits of $\sigma_s(\gamma)$ to $\sigma(\gamma)$. As we found previously in Fig.~\ref{fig7}, $\sigma_m$ decreases with increasing cooling rate and the disordered FCC structures possess the smallest $\sigma_m$. $\gamma_m$ increases weakly with increasing cooling rate and is the largest for the disordered FCC structures. These results illustrate the high-quality fits of the spring network model and emphasize that we can in principle determine the macroscopic stress versus strain relation by determining the {\it local} structural and mechanical properties of metallic glasses. 

\begin{figure*}
\centering
\includegraphics[scale=0.54]{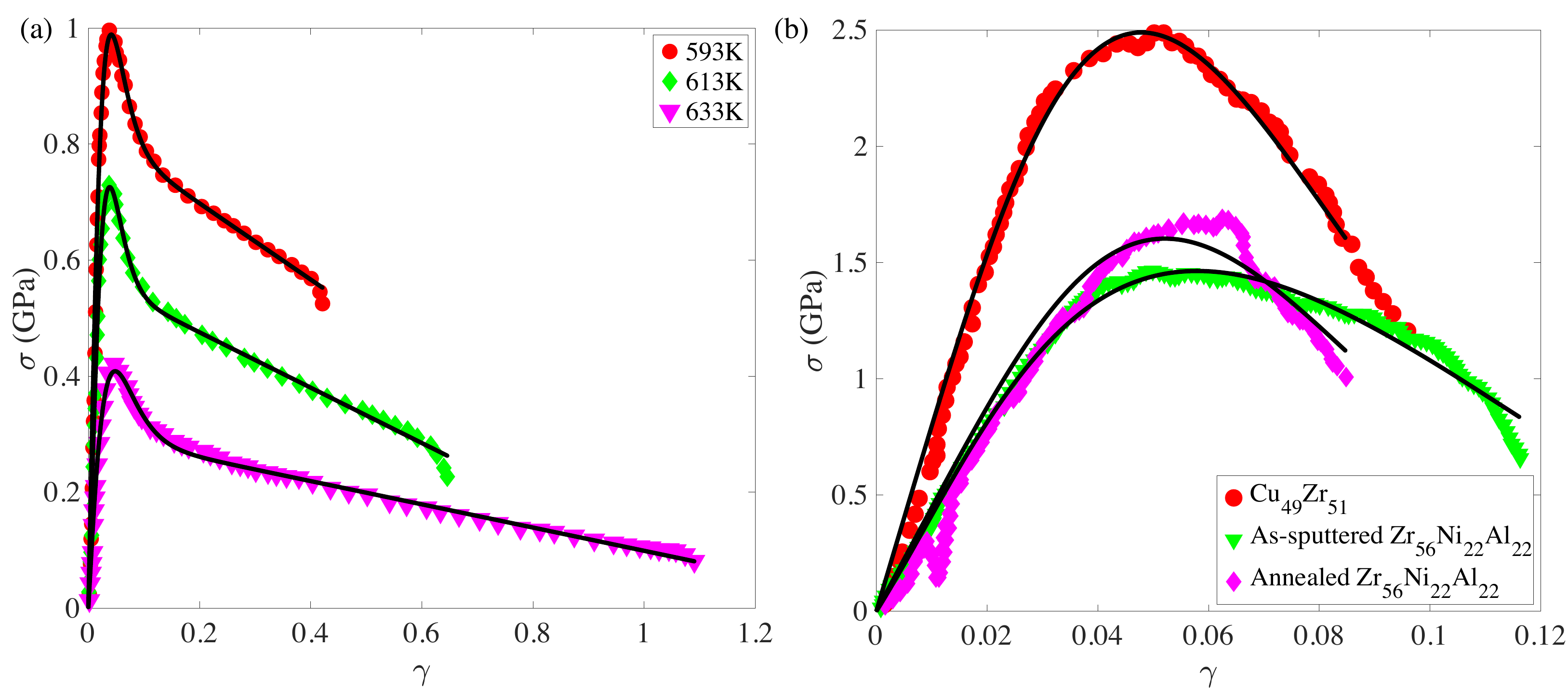}
\caption{Engineering stress $\sigma$ versus strain $\gamma$ from experimental studies of uniaxial tensile deformations applied to several metallic glasses: (a) $\rm Zr_{65}Al_{10}Ni_{10}Cu_{15}$ samples tested at $\rm 593K$ (red circles), $\rm 613K$ (green diamonds), and $\rm 633K$ (magenta triangles)~\cite{RN111} and (b) a nm-scale $\rm Cu_{49}Zr_{51}$~\cite{RN12} sample (red circles) and two sputtered $\rm Zr_{56}Ni_{22}Al_{22}$~\cite{RN25} samples. The $\rm Zr_{56}Ni_{22}Al_{22}$ samples were as sputtered (green diamonds) and annealed below $T_g$ (magenta triangles). The black solid lines are best fits to $\sigma_s(\gamma)$ (Eq.~\ref{predicted_stress}) from the spring network model.}
\label{fig12}
\end{figure*}

Finally, in Fig.~\ref{fig12}, we compare the predictions of $\sigma_s(\gamma)$ from the spring network model to the engineering stress versus strain from experimental studies of uniaxial tensile deformations of ${\rm Zr}_{65}{\rm Al}_{10}{\rm Ni}_{10}{\rm Cu}_{15}$, ${\rm Cu}_{49}{\rm Zr}_{51}$, and ${\rm Zr}_{56}{\rm Ni}_{22}{\rm Al}_{22}$ metallic glass samples~\cite{RN12,RN25,RN111}. In Fig.{~\ref{fig12}} (a), we show $\sigma(\gamma)$ obtained from deforming ${\rm Zr}_{65}{\rm Al}_{10}{\rm Ni}_{10}{\rm Cu}_{15}$ samples with cross-sectional area $0.02$ ${\rm mm^2}$ at strain rate ${\dot \gamma}= 5\times10^{-4} {\rm s}^{-1}$ and tested at temperatures $T=593\rm K$, $613\rm K$, and $633\rm K${~\cite{RN111}}. These prior experimental studies find that $\sigma(\gamma)$ for the ${\rm Zr}_{65}{\rm Al}_{10}{\rm Ni}_{10}{\rm Cu}_{15}$ samples possess large-strain tails prior to fracture for $T >513\rm K$ ({\it cf.} Fig.~\ref{fig7}). $\sigma_s(\gamma)$ from the spring network model can fit $\sigma(\gamma)$ for ${\rm Zr}_{65}{\rm Al}_{10}{\rm Ni}_{10}{\rm Cu}_{15}$ with normalized root-mean-square error values $\langle \Delta \sigma \rangle/\langle \sigma \rangle \lesssim 0.04$. In Fig.~\ref{fig9} (b) and (d), we show the optimal parameters for the spring network model obtained from best fits of $\sigma_s(\gamma)$ to $\sigma(\gamma)$ from the ${\rm Zr}_{65}{\rm Al}_{10}{\rm Ni}_{10}{\rm Cu}_{15}$ sample strained at $T=593\rm K$. The best-fit spring network parameters to the expeirmental results are of the same order of magnitude as those obtained from best-fits to the athermal, quasistatic uniaxial tension simulations of Cu$_{50}$Zr$_{50}$ modeled using the LJ and EAM models.  (Note that EAM potentials are not publicly available for ${\rm Zr}{\rm Al}{\rm Ni}{\rm Cu}$ and ${\rm Zr}{\rm Ni}{\rm Al}$ alloys~\cite{RN113,RN114,RN115}.)

Fig.{~\ref{fig12}} (b) features $\sigma(\gamma)$ obtained from uniaxial tensile deformation of ${\rm Cu}_{49}{\rm Zr}_{51}$ and ${\rm Zr}_{56}{\rm Ni}_{22}{\rm Al}_{22}$ metallic glass samples. The ${\rm Cu}_{49}{\rm Zr}_{51}$ sample has a diameter of $80 {\rm nm}$ and is pulled at strain rate ${\dot \gamma} \sim 10^{-3} {\rm s}^{-1}$ at room temperature~\cite{RN12}. We also considered two sputtered ${\rm Zr}_{56}{\rm Ni}_{22}{\rm Al}_{22}$ samples that have thickness $\sim 90 {\rm nm}$ and are also strained at ${\dot \gamma} \sim 10^{-3}{\rm s}^{-1}$ at room temperature~\cite{RN25}. (As-sputtered samples experience cooling rates in the range $10^8-10^{10} \rm K/s$~\cite{RN47,RN48,RN49}.) One of the two sputtered samples was annealed at sub-$T_g$ temperatures. $\rm ZrNiAl$ alloys have been found to be good glass-formers due to the addition of $\rm Al$~\cite{RN45,RN46}.  The metallic glass samples strained at room temperature do not exhibit large-strain tails in $\sigma(\gamma)$, and instead the samples fracture at small strains $\gamma \sim 0.1$. As a result, the optimal values for the number and rate of change of new springs with strain, $kl_0 N_{n}(0)/A_0$ and $k l_0(dN_n/d\delta\gamma)/A_0$, are close to zero, and the number of initial springs and distribution of spring breaking thresholds $P(\gamma_c)$ determine $\sigma(\gamma)$. Even in this case, the best fits of $\sigma_s(\gamma)$ from the spring network model to $\sigma(\gamma)$ from the experiments give normalized root-mean-square error (rms) values $\langle \Delta \sigma \rangle/\langle \sigma \rangle \lesssim 0.1$. 

\section{Conclusions and Future Directions}
\label{conclusions}

In this article, we developed a novel coarse-grained spring network model to describe the mechanical response of metallic glasses to uniaxial tension. We first performed athermal, quasistatic uniaxial tension simulations of $\rm Cu_{50}Zr_{50}$ metallic glasses modeled using both the Lennard-Jones and EAM potentials. From these simulations, we calculated the engineering stress versus strain $\sigma(\gamma)$ from samples generated over a wide range of cooling rates and different amounts of local positional order. In general, $\sigma(\gamma)$ had the same qualitative form for both the LJ and EAM simulations. We found that the peak $\sigma_m$ in the engineering stress versus strain decreases as the total potential energy per atom of the undeformed structure increases, which shows that one can predict key features of $\sigma(\gamma)$ without actually performing the tensile tests.  Further, we showed that the disordered FCC structures possess more ductile mechanical response than the rapidly cooled metallic glass samples. We analytically solved the one-dimensional spring network model for the total force exerted by the springs as a function of strain.  In the spring network model, initial springs present at $\gamma=0$ stretch during the applied strain and break when they exceed cutoff strains $\gamma_c$ that are selected from a Gamma distribution $P(\gamma_c)$. In addition, new springs can form, and then stretch and break in the same way as the initial springs. 

The engineering stress $\sigma_s(\gamma)$ predicted from the spring network model includes five parameters. Two of the parameters define the shape of the cutoff strain distribution $P(\gamma_c)$. The three other parameters are related to the number of initial springs, and the number of new springs and rate of change of the number of new springs with strain.  We showed that we can express these five parameters in terms of important features of the shape of $\sigma(\gamma)$, i.e., the slope $d\sigma/d\gamma$ at $\gamma=0$, maximum stress $\sigma_m$, strain $\gamma_m$ at which the maximum stress occurs, strain $\gamma_f$ at fracture (i.e. $\sigma=0$ at large strains), and slope $d\sigma/d\gamma$ at $\gamma_f$.  After fitting the spring network model to $\sigma(\gamma)$ from the simulations, we found that the cutoff strain distribution shifts to larger strains for more slowly cooled glasses.  In contrast, the number and rate of change of new springs decreases for slowly cooled glasses.  These two sets of changes offset each other at large strains since $\sigma(\gamma)$ is nearly independent of the sample preparation protocol in this regime. Lastly, we showed that the spring network model can be used to describe the results of experimental studies of uniaxial tensile tests of several Zr-based matallic glasses at temperatures above and below room temperature, where they possess large-strain tails in $\sigma(\gamma)$ and where they fracture at small strains, respectively. 

\begin{figure*}
\centering
\includegraphics[scale=0.52]{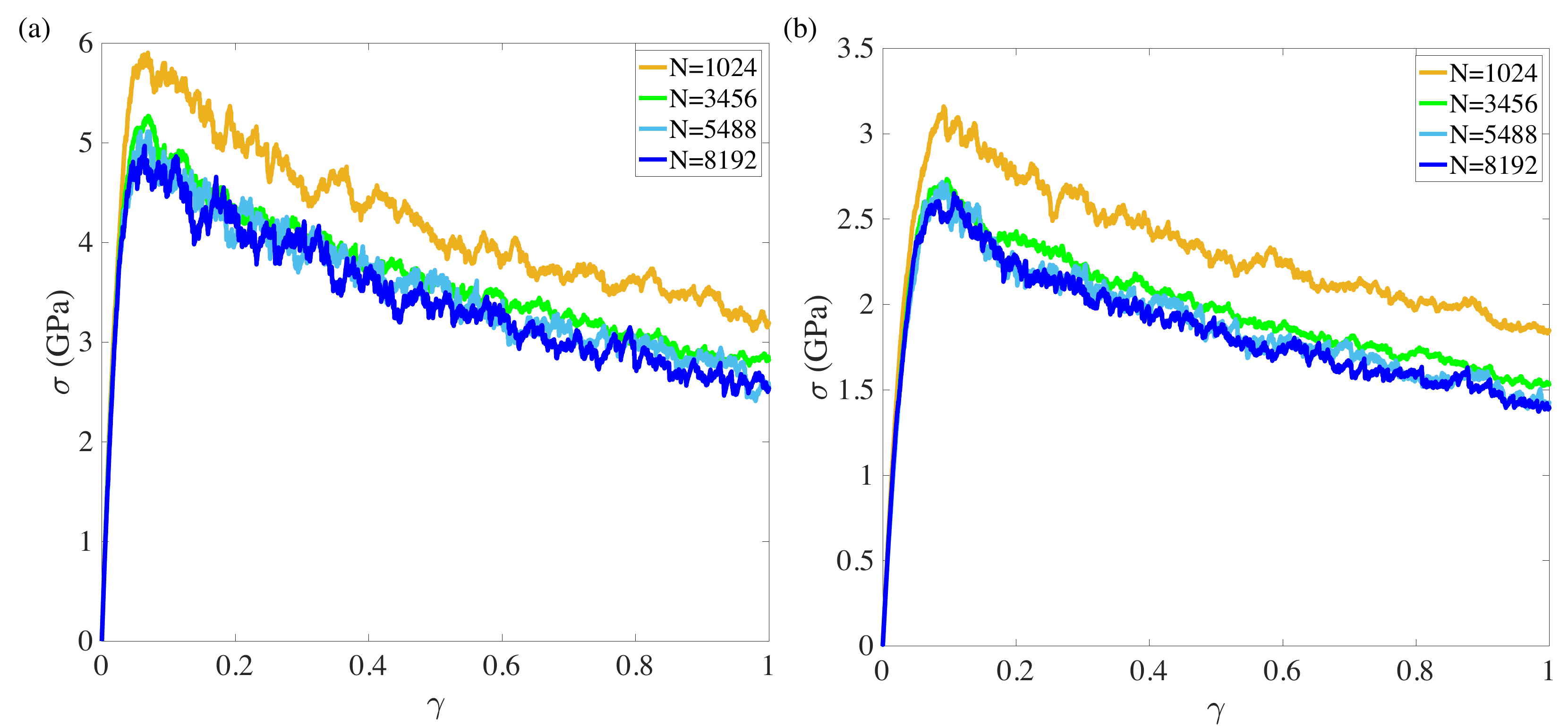}
\caption{Engineering stress $\sigma$ plotted as a function of strain $\gamma$ for athermal, quasistatic simulations of Cu$_{50}$Zr$_{50}$ (modeled using (a) LJ and (b) EAM interactions) undergoing uniaxial strain for system sizes $N = 1024$, $3456$, $5488$, and $8192$. The systems with LJ and EAM interactions were cooled at rates $\rm R=2.6\times10^{14}K/s$ and $\rm 1.0\times10^{14}K/s$, respectively. The engineering stress was averaged over $50$ configurations for $N= 1024$ and $3456$ and over $10$ configurations for the other system sizes.}
\label{fig13}
\end{figure*}

There are several promising directions for future work. First, we will focus on extracting the spring model parameters from the atomistic simulations of uniaxial tension. For example, we can estimate the number of initial springs by determining the distribution of the sizes of the stress drops that occur during uniaxial tension of single, finite-sized samples. To determine the shape of the cutoff strain distribution, number of new springs, and their rate of change with strain from the atomistic simulations, we can add a perturbation to the position of a single atom, minimize the total potential energy, and measure changes in nearest neighbor atoms and the engineering stress as a function of the size of the perturbation. Second, in the current work, we focused on systems that do not form large-scale shear bands.  In future work, we will  consider systems prepared using hybrid Monte Carlo and molecular dynamics simulations that can achieve cooling rates lower than $\rm 10^6 K/s$~\cite{RN93} and display shear banding. We will show that the spring network model can also be used to describe the brittle mechanical response that occurs in systems that form shear bands~\cite{RN110}. Third, we will generalize the one-dimensional spring network model consisting of parallel springs to include springs in series and in parallel. The combination of in series and in parallel springs would enable the modeling of cylcic rapid decreases and increases in stress that occur during uniaxial tension in partially crystalline samples. We can also generalize the spring network model to include springs in series and in parallel in two and three dimensions, which would allow us to model the mechanical response of metallic glasses to simple and pure shear deformations.

\begin{acknowledgments}
The authors acknowledge support from NSF Grant No. CMMI-1901959 (A.N.and C.S.O.). This work was also supported by the High Performance Computing facilities operated by Yale’s Center for Research Computing.
\end{acknowledgments}

\appendix
\section{System-size Effects on Engineering Stress versus Strain}
\label{appendix}

In this appendix, we investigate the dependence of the engineering stress versus strain on system size for the athermal, quasistatic simulations of uniaxial strain. In Fig.~\ref{fig13}, we show $\sigma$ versus $\gamma$ for Cu$_{50}$Zr$_{50}$ metallic glass samples modeled using (a) LJ and (b) EAM interactions for several system sizes, $N=1024$, $3456$, $5488$, and $8192$.  The systems with LJ and EAM interactions were prepared using cooling rates $R=\rm 2.6\times10^{14}K/s$ and $\rm 1.0\times10^{14}K/s$, respectively.  (We find results similar to those in Fig.~\ref{fig13} for the other cooling rates considered in the main text.) For systems with both LJ and EAM interactions, $\sigma(\gamma)$ for $N=1024$ differs significantly from the other system sizes, whereas $\sigma(\gamma)$ for $N \ge 3456$ does not depend on system size. Thus, in the main text we present results on $\sigma(\gamma)$ for systems with $N=3456$ atoms.  

\bibliography{ref}

\begin{thebibliography}{60}%
\makeatletter
\providecommand \@ifxundefined [1]{%
 \@ifx{#1\undefined}
}%
\providecommand \@ifnum [1]{%
 \ifnum #1\expandafter \@firstoftwo
 \else \expandafter \@secondoftwo
 \fi
}%
\providecommand \@ifx [1]{%
 \ifx #1\expandafter \@firstoftwo
 \else \expandafter \@secondoftwo
 \fi
}%
\providecommand \natexlab [1]{#1}%
\providecommand \enquote  [1]{``#1''}%
\providecommand \bibnamefont  [1]{#1}%
\providecommand \bibfnamefont [1]{#1}%
\providecommand \citenamefont [1]{#1}%
\providecommand \href@noop [0]{\@secondoftwo}%
\providecommand \href [0]{\begingroup \@sanitize@url \@href}%
\providecommand \@href[1]{\@@startlink{#1}\@@href}%
\providecommand \@@href[1]{\endgroup#1\@@endlink}%
\providecommand \@sanitize@url [0]{\catcode `\\12\catcode `\$12\catcode
  `\&12\catcode `\#12\catcode `\^12\catcode `\_12\catcode `\%12\relax}%
\providecommand \@@startlink[1]{}%
\providecommand \@@endlink[0]{}%
\providecommand \url  [0]{\begingroup\@sanitize@url \@url }%
\providecommand \@url [1]{\endgroup\@href {#1}{\urlprefix }}%
\providecommand \urlprefix  [0]{URL }%
\providecommand \Eprint [0]{\href }%
\providecommand \doibase [0]{https://doi.org/}%
\providecommand \selectlanguage [0]{\@gobble}%
\providecommand \bibinfo  [0]{\@secondoftwo}%
\providecommand \bibfield  [0]{\@secondoftwo}%
\providecommand \translation [1]{[#1]}%
\providecommand \BibitemOpen [0]{}%
\providecommand \bibitemStop [0]{}%
\providecommand \bibitemNoStop [0]{.\EOS\space}%
\providecommand \EOS [0]{\spacefactor3000\relax}%
\providecommand \BibitemShut  [1]{\csname bibitem#1\endcsname}%
\let\auto@bib@innerbib\@empty
\bibitem [{\citenamefont {Ashby}\ and\ \citenamefont {Greer}(2006)}]{RN65}%
  \BibitemOpen
  \bibfield  {author} {\bibinfo {author} {\bibfnamefont {M.~F.}\ \bibnamefont
  {Ashby}}\ and\ \bibinfo {author} {\bibfnamefont {A.~L.}\ \bibnamefont
  {Greer}},\ }\href
  {https://doi.org/https://doi.org/10.1016/j.scriptamat.2005.09.051} {\bibfield
   {journal} {\bibinfo  {journal} {Scripta Materialia}\ }\textbf {\bibinfo
  {volume} {54}},\ \bibinfo {pages} {321} (\bibinfo {year} {2006})}\BibitemShut
  {NoStop}%
\bibitem [{\citenamefont {Schuh}\ \emph {et~al.}(2007)\citenamefont {Schuh},
  \citenamefont {Hufnagel},\ and\ \citenamefont {Ramamurty}}]{RN66}%
  \BibitemOpen
  \bibfield  {author} {\bibinfo {author} {\bibfnamefont {C.~A.}\ \bibnamefont
  {Schuh}}, \bibinfo {author} {\bibfnamefont {T.~C.}\ \bibnamefont
  {Hufnagel}},\ and\ \bibinfo {author} {\bibfnamefont {U.}~\bibnamefont
  {Ramamurty}},\ }\href
  {https://doi.org/https://doi.org/10.1016/j.actamat.2007.01.052} {\bibfield
  {journal} {\bibinfo  {journal} {Acta Materialia}\ }\textbf {\bibinfo {volume}
  {55}},\ \bibinfo {pages} {4067} (\bibinfo {year} {2007})}\BibitemShut
  {NoStop}%
\bibitem [{\citenamefont {Liontas}\ \emph {et~al.}(2016)\citenamefont
  {Liontas}, \citenamefont {Jafary-Zadeh}, \citenamefont {Zeng}, \citenamefont
  {Zhang}, \citenamefont {Mao},\ and\ \citenamefont {Greer}}]{RN25}%
  \BibitemOpen
  \bibfield  {author} {\bibinfo {author} {\bibfnamefont {R.}~\bibnamefont
  {Liontas}}, \bibinfo {author} {\bibfnamefont {M.}~\bibnamefont
  {Jafary-Zadeh}}, \bibinfo {author} {\bibfnamefont {Q.}~\bibnamefont {Zeng}},
  \bibinfo {author} {\bibfnamefont {Y.-W.}\ \bibnamefont {Zhang}}, \bibinfo
  {author} {\bibfnamefont {W.~L.}\ \bibnamefont {Mao}},\ and\ \bibinfo {author}
  {\bibfnamefont {J.~R.}\ \bibnamefont {Greer}},\ }\href
  {https://doi.org/https://doi.org/10.1016/j.actamat.2016.07.050} {\bibfield
  {journal} {\bibinfo  {journal} {Acta Materialia}\ }\textbf {\bibinfo {volume}
  {118}},\ \bibinfo {pages} {270} (\bibinfo {year} {2016})}\BibitemShut
  {NoStop}%
\bibitem [{\citenamefont {Magagnosc}\ \emph {et~al.}(2013)\citenamefont
  {Magagnosc}, \citenamefont {Ehrbar}, \citenamefont {Kumar}, \citenamefont
  {He}, \citenamefont {Schroers},\ and\ \citenamefont {Gianola}}]{RN109}%
  \BibitemOpen
  \bibfield  {author} {\bibinfo {author} {\bibfnamefont {D.~J.}\ \bibnamefont
  {Magagnosc}}, \bibinfo {author} {\bibfnamefont {R.}~\bibnamefont {Ehrbar}},
  \bibinfo {author} {\bibfnamefont {G.}~\bibnamefont {Kumar}}, \bibinfo
  {author} {\bibfnamefont {M.~R.}\ \bibnamefont {He}}, \bibinfo {author}
  {\bibfnamefont {J.}~\bibnamefont {Schroers}},\ and\ \bibinfo {author}
  {\bibfnamefont {D.~S.}\ \bibnamefont {Gianola}},\ }\href
  {https://doi.org/10.1038/srep01096} {\bibfield  {journal} {\bibinfo
  {journal} {Scientific Reports}\ }\textbf {\bibinfo {volume} {3}},\ \bibinfo
  {pages} {1096} (\bibinfo {year} {2013})}\BibitemShut {NoStop}%
\bibitem [{\citenamefont {Guo}\ \emph {et~al.}(2007)\citenamefont {Guo},
  \citenamefont {Yan}, \citenamefont {Wang}, \citenamefont {Tan}, \citenamefont
  {Zhang}, \citenamefont {Sui},\ and\ \citenamefont {Ma}}]{RN15}%
  \BibitemOpen
  \bibfield  {author} {\bibinfo {author} {\bibfnamefont {H.}~\bibnamefont
  {Guo}}, \bibinfo {author} {\bibfnamefont {P.~F.}\ \bibnamefont {Yan}},
  \bibinfo {author} {\bibfnamefont {Y.~B.}\ \bibnamefont {Wang}}, \bibinfo
  {author} {\bibfnamefont {J.}~\bibnamefont {Tan}}, \bibinfo {author}
  {\bibfnamefont {Z.~F.}\ \bibnamefont {Zhang}}, \bibinfo {author}
  {\bibfnamefont {M.~L.}\ \bibnamefont {Sui}},\ and\ \bibinfo {author}
  {\bibfnamefont {E.}~\bibnamefont {Ma}},\ }\href
  {https://doi.org/10.1038/nmat1984} {\bibfield  {journal} {\bibinfo  {journal}
  {Nature Materials}\ }\textbf {\bibinfo {volume} {6}},\ \bibinfo {pages} {735}
  (\bibinfo {year} {2007})}\BibitemShut {NoStop}%
\bibitem [{\citenamefont {Tian}\ \emph {et~al.}(2013)\citenamefont {Tian},
  \citenamefont {Shan},\ and\ \citenamefont {Ma}}]{RN12}%
  \BibitemOpen
  \bibfield  {author} {\bibinfo {author} {\bibfnamefont {L.}~\bibnamefont
  {Tian}}, \bibinfo {author} {\bibfnamefont {Z.-W.}\ \bibnamefont {Shan}},\
  and\ \bibinfo {author} {\bibfnamefont {E.}~\bibnamefont {Ma}},\ }\href
  {https://doi.org/https://doi.org/10.1016/j.actamat.2013.05.001} {\bibfield
  {journal} {\bibinfo  {journal} {Acta Materialia}\ }\textbf {\bibinfo {volume}
  {61}},\ \bibinfo {pages} {4823} (\bibinfo {year} {2013})}\BibitemShut
  {NoStop}%
\bibitem [{\citenamefont {Yi}\ \emph {et~al.}(2015)\citenamefont {Yi},
  \citenamefont {Wang},\ and\ \citenamefont {Lewandowski}}]{RN24}%
  \BibitemOpen
  \bibfield  {author} {\bibinfo {author} {\bibfnamefont {J.}~\bibnamefont
  {Yi}}, \bibinfo {author} {\bibfnamefont {W.~H.}\ \bibnamefont {Wang}},\ and\
  \bibinfo {author} {\bibfnamefont {J.~J.}\ \bibnamefont {Lewandowski}},\
  }\href {https://doi.org/https://doi.org/10.1016/j.actamat.2014.12.039}
  {\bibfield  {journal} {\bibinfo  {journal} {Acta Materialia}\ }\textbf
  {\bibinfo {volume} {87}},\ \bibinfo {pages} {1} (\bibinfo {year}
  {2015})}\BibitemShut {NoStop}%
\bibitem [{\citenamefont {Yokoyama}\ \emph {et~al.}(2009)\citenamefont
  {Yokoyama}, \citenamefont {Fujita}, \citenamefont {Yavari},\ and\
  \citenamefont {Inoue}}]{RN9}%
  \BibitemOpen
  \bibfield  {author} {\bibinfo {author} {\bibfnamefont {Y.}~\bibnamefont
  {Yokoyama}}, \bibinfo {author} {\bibfnamefont {K.}~\bibnamefont {Fujita}},
  \bibinfo {author} {\bibfnamefont {A.~R.}\ \bibnamefont {Yavari}},\ and\
  \bibinfo {author} {\bibfnamefont {A.}~\bibnamefont {Inoue}},\ }\href
  {https://doi.org/10.1080/09500830902873575} {\bibfield  {journal} {\bibinfo
  {journal} {Philosophical Magazine Letters}\ }\textbf {\bibinfo {volume}
  {89}},\ \bibinfo {pages} {322} (\bibinfo {year} {2009})}\BibitemShut
  {NoStop}%
\bibitem [{\citenamefont {Yu}\ \emph {et~al.}(2012)\citenamefont {Yu},
  \citenamefont {Shen}, \citenamefont {Wang}, \citenamefont {Gu}, \citenamefont
  {Wang},\ and\ \citenamefont {Bai}}]{RN10}%
  \BibitemOpen
  \bibfield  {author} {\bibinfo {author} {\bibfnamefont {H.~B.}\ \bibnamefont
  {Yu}}, \bibinfo {author} {\bibfnamefont {X.}~\bibnamefont {Shen}}, \bibinfo
  {author} {\bibfnamefont {Z.}~\bibnamefont {Wang}}, \bibinfo {author}
  {\bibfnamefont {L.}~\bibnamefont {Gu}}, \bibinfo {author} {\bibfnamefont
  {W.~H.}\ \bibnamefont {Wang}},\ and\ \bibinfo {author} {\bibfnamefont
  {H.~Y.}\ \bibnamefont {Bai}},\ }\href
  {https://doi.org/10.1103/PhysRevLett.108.015504} {\bibfield  {journal}
  {\bibinfo  {journal} {Physical Review Letters}\ }\textbf {\bibinfo {volume}
  {108}},\ \bibinfo {pages} {015504} (\bibinfo {year} {2012})}\BibitemShut
  {NoStop}%
\bibitem [{\citenamefont {Kawamura}\ \emph {et~al.}(1997)\citenamefont
  {Kawamura}, \citenamefont {Shibata}, \citenamefont {Inoue},\ and\
  \citenamefont {Masumoto}}]{RN111}%
  \BibitemOpen
  \bibfield  {author} {\bibinfo {author} {\bibfnamefont {Y.}~\bibnamefont
  {Kawamura}}, \bibinfo {author} {\bibfnamefont {T.}~\bibnamefont {Shibata}},
  \bibinfo {author} {\bibfnamefont {A.}~\bibnamefont {Inoue}},\ and\ \bibinfo
  {author} {\bibfnamefont {T.}~\bibnamefont {Masumoto}},\ }\href
  {https://doi.org/https://doi.org/10.1016/S1359-6462(97)00105-X} {\bibfield
  {journal} {\bibinfo  {journal} {Scripta Materialia}\ }\textbf {\bibinfo
  {volume} {37}},\ \bibinfo {pages} {431} (\bibinfo {year} {1997})}\BibitemShut
  {NoStop}%
\bibitem [{\citenamefont {Kumar}\ \emph {et~al.}(2011)\citenamefont {Kumar},
  \citenamefont {Prades-Rodel}, \citenamefont {Blatter},\ and\ \citenamefont
  {Schroers}}]{RN67}%
  \BibitemOpen
  \bibfield  {author} {\bibinfo {author} {\bibfnamefont {G.}~\bibnamefont
  {Kumar}}, \bibinfo {author} {\bibfnamefont {S.}~\bibnamefont {Prades-Rodel}},
  \bibinfo {author} {\bibfnamefont {A.}~\bibnamefont {Blatter}},\ and\ \bibinfo
  {author} {\bibfnamefont {J.}~\bibnamefont {Schroers}},\ }\href
  {https://doi.org/https://doi.org/10.1016/j.scriptamat.2011.06.029} {\bibfield
   {journal} {\bibinfo  {journal} {Scripta Materialia}\ }\textbf {\bibinfo
  {volume} {65}},\ \bibinfo {pages} {585} (\bibinfo {year} {2011})}\BibitemShut
  {NoStop}%
\bibitem [{\citenamefont {Kumar}\ \emph {et~al.}(2007)\citenamefont {Kumar},
  \citenamefont {Ohkubo}, \citenamefont {Mukai},\ and\ \citenamefont
  {Hono}}]{RN68}%
  \BibitemOpen
  \bibfield  {author} {\bibinfo {author} {\bibfnamefont {G.}~\bibnamefont
  {Kumar}}, \bibinfo {author} {\bibfnamefont {T.}~\bibnamefont {Ohkubo}},
  \bibinfo {author} {\bibfnamefont {T.}~\bibnamefont {Mukai}},\ and\ \bibinfo
  {author} {\bibfnamefont {K.}~\bibnamefont {Hono}},\ }\href
  {https://doi.org/https://doi.org/10.1016/j.scriptamat.2007.02.013} {\bibfield
   {journal} {\bibinfo  {journal} {Scripta Materialia}\ }\textbf {\bibinfo
  {volume} {57}},\ \bibinfo {pages} {173} (\bibinfo {year} {2007})}\BibitemShut
  {NoStop}%
\bibitem [{\citenamefont {Das}\ \emph {et~al.}(2005)\citenamefont {Das},
  \citenamefont {Tang}, \citenamefont {Kim}, \citenamefont {Theissmann},
  \citenamefont {Baier}, \citenamefont {Wang},\ and\ \citenamefont
  {Eckert}}]{RN69}%
  \BibitemOpen
  \bibfield  {author} {\bibinfo {author} {\bibfnamefont {J.}~\bibnamefont
  {Das}}, \bibinfo {author} {\bibfnamefont {M.~B.}\ \bibnamefont {Tang}},
  \bibinfo {author} {\bibfnamefont {K.~B.}\ \bibnamefont {Kim}}, \bibinfo
  {author} {\bibfnamefont {R.}~\bibnamefont {Theissmann}}, \bibinfo {author}
  {\bibfnamefont {F.}~\bibnamefont {Baier}}, \bibinfo {author} {\bibfnamefont
  {W.~H.}\ \bibnamefont {Wang}},\ and\ \bibinfo {author} {\bibfnamefont
  {J.}~\bibnamefont {Eckert}},\ }\href
  {https://doi.org/10.1103/PhysRevLett.94.205501} {\bibfield  {journal}
  {\bibinfo  {journal} {Physical Review Letters}\ }\textbf {\bibinfo {volume}
  {94}},\ \bibinfo {pages} {205501} (\bibinfo {year} {2005})}\BibitemShut
  {NoStop}%
\bibitem [{\citenamefont {Nicolas}\ \emph {et~al.}(2018)\citenamefont
  {Nicolas}, \citenamefont {Ferrero}, \citenamefont {Martens},\ and\
  \citenamefont {Barrat}}]{RN8}%
  \BibitemOpen
  \bibfield  {author} {\bibinfo {author} {\bibfnamefont {A.}~\bibnamefont
  {Nicolas}}, \bibinfo {author} {\bibfnamefont {E.~E.}\ \bibnamefont
  {Ferrero}}, \bibinfo {author} {\bibfnamefont {K.}~\bibnamefont {Martens}},\
  and\ \bibinfo {author} {\bibfnamefont {J.-L.}\ \bibnamefont {Barrat}},\
  }\href {https://doi.org/10.1103/RevModPhys.90.045006} {\bibfield  {journal}
  {\bibinfo  {journal} {Reviews of Modern Physics}\ }\textbf {\bibinfo {volume}
  {90}},\ \bibinfo {pages} {045006} (\bibinfo {year} {2018})}\BibitemShut
  {NoStop}%
\bibitem [{\citenamefont {Homer}\ and\ \citenamefont {Schuh}(2009)}]{RN2}%
  \BibitemOpen
  \bibfield  {author} {\bibinfo {author} {\bibfnamefont {E.~R.}\ \bibnamefont
  {Homer}}\ and\ \bibinfo {author} {\bibfnamefont {C.~A.}\ \bibnamefont
  {Schuh}},\ }\href
  {https://doi.org/https://doi.org/10.1016/j.actamat.2009.02.035} {\bibfield
  {journal} {\bibinfo  {journal} {Acta Materialia}\ }\textbf {\bibinfo {volume}
  {57}},\ \bibinfo {pages} {2823} (\bibinfo {year} {2009})}\BibitemShut
  {NoStop}%
\bibitem [{\citenamefont {Li}\ \emph {et~al.}(2013)\citenamefont {Li},
  \citenamefont {Homer},\ and\ \citenamefont {Schuh}}]{RN1}%
  \BibitemOpen
  \bibfield  {author} {\bibinfo {author} {\bibfnamefont {L.}~\bibnamefont
  {Li}}, \bibinfo {author} {\bibfnamefont {E.~R.}\ \bibnamefont {Homer}},\ and\
  \bibinfo {author} {\bibfnamefont {C.~A.}\ \bibnamefont {Schuh}},\ }\href
  {https://doi.org/https://doi.org/10.1016/j.actamat.2013.02.024} {\bibfield
  {journal} {\bibinfo  {journal} {Acta Materialia}\ }\textbf {\bibinfo {volume}
  {61}},\ \bibinfo {pages} {3347} (\bibinfo {year} {2013})}\BibitemShut
  {NoStop}%
\bibitem [{\citenamefont {Zhang}\ \emph
  {et~al.}(2022{\natexlab{a}})\citenamefont {Zhang}, \citenamefont {Xiao},
  \citenamefont {Yang}, \citenamefont {Ivancic}, \citenamefont {Ridout},
  \citenamefont {Riggleman}, \citenamefont {Durian},\ and\ \citenamefont
  {Liu}}]{RN81}%
  \BibitemOpen
  \bibfield  {author} {\bibinfo {author} {\bibfnamefont {G.}~\bibnamefont
  {Zhang}}, \bibinfo {author} {\bibfnamefont {H.}~\bibnamefont {Xiao}},
  \bibinfo {author} {\bibfnamefont {E.}~\bibnamefont {Yang}}, \bibinfo {author}
  {\bibfnamefont {R.~J.~S.}\ \bibnamefont {Ivancic}}, \bibinfo {author}
  {\bibfnamefont {S.~A.}\ \bibnamefont {Ridout}}, \bibinfo {author}
  {\bibfnamefont {R.~A.}\ \bibnamefont {Riggleman}}, \bibinfo {author}
  {\bibfnamefont {D.~J.}\ \bibnamefont {Durian}},\ and\ \bibinfo {author}
  {\bibfnamefont {A.~J.}\ \bibnamefont {Liu}},\ }\href
  {https://doi.org/10.1103/PhysRevResearch.4.043026} {\bibfield  {journal}
  {\bibinfo  {journal} {Physical Review Research}\ }\textbf {\bibinfo {volume}
  {4}},\ \bibinfo {pages} {043026} (\bibinfo {year}
  {2022}{\natexlab{a}})}\BibitemShut {NoStop}%
\bibitem [{\citenamefont {Liu}\ \emph {et~al.}(2021{\natexlab{a}})\citenamefont
  {Liu}, \citenamefont {Dutta}, \citenamefont {Chaudhuri},\ and\ \citenamefont
  {Martens}}]{RN34}%
  \BibitemOpen
  \bibfield  {author} {\bibinfo {author} {\bibfnamefont {C.}~\bibnamefont
  {Liu}}, \bibinfo {author} {\bibfnamefont {S.}~\bibnamefont {Dutta}}, \bibinfo
  {author} {\bibfnamefont {P.}~\bibnamefont {Chaudhuri}},\ and\ \bibinfo
  {author} {\bibfnamefont {K.}~\bibnamefont {Martens}},\ }\href
  {https://doi.org/10.1103/PhysRevLett.126.138005} {\bibfield  {journal}
  {\bibinfo  {journal} {Physical Review Letters}\ }\textbf {\bibinfo {volume}
  {126}},\ \bibinfo {pages} {138005} (\bibinfo {year}
  {2021}{\natexlab{a}})}\BibitemShut {NoStop}%
\bibitem [{\citenamefont {Castellanos}\ \emph {et~al.}(2022)\citenamefont
  {Castellanos}, \citenamefont {Roux},\ and\ \citenamefont {Patinet}}]{RN77}%
  \BibitemOpen
  \bibfield  {author} {\bibinfo {author} {\bibfnamefont {D.~F.}\ \bibnamefont
  {Castellanos}}, \bibinfo {author} {\bibfnamefont {S.}~\bibnamefont {Roux}},\
  and\ \bibinfo {author} {\bibfnamefont {S.}~\bibnamefont {Patinet}},\ }\href
  {https://doi.org/https://doi.org/10.1016/j.actamat.2022.118405} {\bibfield
  {journal} {\bibinfo  {journal} {Acta Materialia}\ }\textbf {\bibinfo {volume}
  {241}},\ \bibinfo {pages} {118405} (\bibinfo {year} {2022})}\BibitemShut
  {NoStop}%
\bibitem [{\citenamefont {Falk}\ and\ \citenamefont {Langer}(2011)}]{RN90}%
  \BibitemOpen
  \bibfield  {author} {\bibinfo {author} {\bibfnamefont {M.~L.}\ \bibnamefont
  {Falk}}\ and\ \bibinfo {author} {\bibfnamefont {J.~S.}\ \bibnamefont
  {Langer}},\ }\href {https://doi.org/10.1146/annurev-conmatphys-062910-140452}
  {\bibfield  {journal} {\bibinfo  {journal} {Annual Review of Condensed Matter
  Physics}\ }\textbf {\bibinfo {volume} {2}},\ \bibinfo {pages} {353} (\bibinfo
  {year} {2011})}\BibitemShut {NoStop}%
\bibitem [{\citenamefont {Manning}\ \emph {et~al.}(2007)\citenamefont
  {Manning}, \citenamefont {Langer},\ and\ \citenamefont {Carlson}}]{RN79}%
  \BibitemOpen
  \bibfield  {author} {\bibinfo {author} {\bibfnamefont {M.~L.}\ \bibnamefont
  {Manning}}, \bibinfo {author} {\bibfnamefont {J.~S.}\ \bibnamefont
  {Langer}},\ and\ \bibinfo {author} {\bibfnamefont {J.~M.}\ \bibnamefont
  {Carlson}},\ }\href {https://doi.org/10.1103/PhysRevE.76.056106} {\bibfield
  {journal} {\bibinfo  {journal} {Physical Review E}\ }\textbf {\bibinfo
  {volume} {76}},\ \bibinfo {pages} {056106} (\bibinfo {year}
  {2007})}\BibitemShut {NoStop}%
\bibitem [{\citenamefont {Hinkle}\ \emph {et~al.}(2017)\citenamefont {Hinkle},
  \citenamefont {Rycroft}, \citenamefont {Shields},\ and\ \citenamefont
  {Falk}}]{RN33}%
  \BibitemOpen
  \bibfield  {author} {\bibinfo {author} {\bibfnamefont {A.~R.}\ \bibnamefont
  {Hinkle}}, \bibinfo {author} {\bibfnamefont {C.~H.}\ \bibnamefont {Rycroft}},
  \bibinfo {author} {\bibfnamefont {M.~D.}\ \bibnamefont {Shields}},\ and\
  \bibinfo {author} {\bibfnamefont {M.~L.}\ \bibnamefont {Falk}},\ }\href
  {https://doi.org/10.1103/PhysRevE.95.053001} {\bibfield  {journal} {\bibinfo
  {journal} {Physical Review E}\ }\textbf {\bibinfo {volume} {95}},\ \bibinfo
  {pages} {053001} (\bibinfo {year} {2017})}\BibitemShut {NoStop}%
\bibitem [{\citenamefont {Alava}\ \emph {et~al.}(2006)\citenamefont {Alava},
  \citenamefont {Nukala},\ and\ \citenamefont {Zapperi}}]{RN4}%
  \BibitemOpen
  \bibfield  {author} {\bibinfo {author} {\bibfnamefont {M.~J.}\ \bibnamefont
  {Alava}}, \bibinfo {author} {\bibfnamefont {P.~K. V.~V.}\ \bibnamefont
  {Nukala}},\ and\ \bibinfo {author} {\bibfnamefont {S.}~\bibnamefont
  {Zapperi}},\ }\href {https://doi.org/10.1080/00018730300741518} {\bibfield
  {journal} {\bibinfo  {journal} {Advances in Physics}\ }\textbf {\bibinfo
  {volume} {55}},\ \bibinfo {pages} {349} (\bibinfo {year} {2006})}\BibitemShut
  {NoStop}%
\bibitem [{\citenamefont {Hemmer}\ and\ \citenamefont {Hansen}(1992)}]{RN5}%
  \BibitemOpen
  \bibfield  {author} {\bibinfo {author} {\bibfnamefont {P.~C.}\ \bibnamefont
  {Hemmer}}\ and\ \bibinfo {author} {\bibfnamefont {A.}~\bibnamefont
  {Hansen}},\ }\href {https://doi.org/10.1115/1.2894060} {\bibfield  {journal}
  {\bibinfo  {journal} {Journal of Applied Mechanics}\ }\textbf {\bibinfo
  {volume} {59}},\ \bibinfo {pages} {909} (\bibinfo {year} {1992})}\BibitemShut
  {NoStop}%
\bibitem [{\citenamefont {Pradhan}\ \emph {et~al.}(2010)\citenamefont
  {Pradhan}, \citenamefont {Hansen},\ and\ \citenamefont {Chakrabarti}}]{RN6}%
  \BibitemOpen
  \bibfield  {author} {\bibinfo {author} {\bibfnamefont {S.}~\bibnamefont
  {Pradhan}}, \bibinfo {author} {\bibfnamefont {A.}~\bibnamefont {Hansen}},\
  and\ \bibinfo {author} {\bibfnamefont {B.~K.}\ \bibnamefont {Chakrabarti}},\
  }\href {https://doi.org/10.1103/RevModPhys.82.499} {\bibfield  {journal}
  {\bibinfo  {journal} {Reviews of Modern Physics}\ }\textbf {\bibinfo {volume}
  {82}},\ \bibinfo {pages} {499} (\bibinfo {year} {2010})}\BibitemShut
  {NoStop}%
\bibitem [{\citenamefont {Roy}\ \emph {et~al.}(2017)\citenamefont {Roy},
  \citenamefont {Biswas},\ and\ \citenamefont {Ray}}]{RN72}%
  \BibitemOpen
  \bibfield  {author} {\bibinfo {author} {\bibfnamefont {S.}~\bibnamefont
  {Roy}}, \bibinfo {author} {\bibfnamefont {S.}~\bibnamefont {Biswas}},\ and\
  \bibinfo {author} {\bibfnamefont {P.}~\bibnamefont {Ray}},\ }\href
  {https://doi.org/10.1103/PhysRevE.96.063003} {\bibfield  {journal} {\bibinfo
  {journal} {Physical Review E}\ }\textbf {\bibinfo {volume} {96}},\ \bibinfo
  {pages} {063003} (\bibinfo {year} {2017})}\BibitemShut {NoStop}%
\bibitem [{\citenamefont {Biswas}\ \emph {et~al.}(2015)\citenamefont {Biswas},
  \citenamefont {Roy},\ and\ \citenamefont {Ray}}]{RN73}%
  \BibitemOpen
  \bibfield  {author} {\bibinfo {author} {\bibfnamefont {S.}~\bibnamefont
  {Biswas}}, \bibinfo {author} {\bibfnamefont {S.}~\bibnamefont {Roy}},\ and\
  \bibinfo {author} {\bibfnamefont {P.}~\bibnamefont {Ray}},\ }\href
  {https://doi.org/10.1103/PhysRevE.91.050105} {\bibfield  {journal} {\bibinfo
  {journal} {Physical Review E}\ }\textbf {\bibinfo {volume} {91}},\ \bibinfo
  {pages} {050105(R)} (\bibinfo {year} {2015})}\BibitemShut {NoStop}%
\bibitem [{\citenamefont {Li}\ \emph {et~al.}(2008)\citenamefont {Li},
  \citenamefont {Guo}, \citenamefont {Kalb},\ and\ \citenamefont
  {Thompson}}]{RN57}%
  \BibitemOpen
  \bibfield  {author} {\bibinfo {author} {\bibfnamefont {Y.}~\bibnamefont
  {Li}}, \bibinfo {author} {\bibfnamefont {Q.}~\bibnamefont {Guo}}, \bibinfo
  {author} {\bibfnamefont {J.~A.}\ \bibnamefont {Kalb}},\ and\ \bibinfo
  {author} {\bibfnamefont {C.~V.}\ \bibnamefont {Thompson}},\ }\href
  {https://doi.org/10.1126/science.1163062} {\bibfield  {journal} {\bibinfo
  {journal} {Science}\ }\textbf {\bibinfo {volume} {322}},\ \bibinfo {pages}
  {1816} (\bibinfo {year} {2008})}\BibitemShut {NoStop}%
\bibitem [{\citenamefont {Wang}\ \emph {et~al.}(2005)\citenamefont {Wang},
  \citenamefont {Lewandowski},\ and\ \citenamefont {Greer}}]{RN42}%
  \BibitemOpen
  \bibfield  {author} {\bibinfo {author} {\bibfnamefont {W.~H.}\ \bibnamefont
  {Wang}}, \bibinfo {author} {\bibfnamefont {J.~J.}\ \bibnamefont
  {Lewandowski}},\ and\ \bibinfo {author} {\bibfnamefont {A.~L.}\ \bibnamefont
  {Greer}},\ }\href {https://doi.org/10.1557/jmr.2005.0302} {\bibfield
  {journal} {\bibinfo  {journal} {Journal of Materials Research}\ }\textbf
  {\bibinfo {volume} {20}},\ \bibinfo {pages} {2307} (\bibinfo {year}
  {2005})}\BibitemShut {NoStop}%
\bibitem [{\citenamefont {Jin}\ \emph {et~al.}(2021)\citenamefont {Jin},
  \citenamefont {Datye}, \citenamefont {Schwarz}, \citenamefont {Shattuck},\
  and\ \citenamefont {O'Hern}}]{RN91}%
  \BibitemOpen
  \bibfield  {author} {\bibinfo {author} {\bibfnamefont {W.}~\bibnamefont
  {Jin}}, \bibinfo {author} {\bibfnamefont {A.}~\bibnamefont {Datye}}, \bibinfo
  {author} {\bibfnamefont {U.~D.}\ \bibnamefont {Schwarz}}, \bibinfo {author}
  {\bibfnamefont {M.~D.}\ \bibnamefont {Shattuck}},\ and\ \bibinfo {author}
  {\bibfnamefont {C.~S.}\ \bibnamefont {O'Hern}},\ }\href
  {https://doi.org/10.1039/D1SM00898F} {\bibfield  {journal} {\bibinfo
  {journal} {Soft Matter}\ }\textbf {\bibinfo {volume} {17}},\ \bibinfo {pages}
  {8612} (\bibinfo {year} {2021})}\BibitemShut {NoStop}%
\bibitem [{\citenamefont {Mendelev}\ \emph {et~al.}(2019)\citenamefont
  {Mendelev}, \citenamefont {Sun}, \citenamefont {Zhang}, \citenamefont
  {Wang},\ and\ \citenamefont {Ho}}]{RN13}%
  \BibitemOpen
  \bibfield  {author} {\bibinfo {author} {\bibfnamefont {M.~I.}\ \bibnamefont
  {Mendelev}}, \bibinfo {author} {\bibfnamefont {Y.}~\bibnamefont {Sun}},
  \bibinfo {author} {\bibfnamefont {F.}~\bibnamefont {Zhang}}, \bibinfo
  {author} {\bibfnamefont {C.~Z.}\ \bibnamefont {Wang}},\ and\ \bibinfo
  {author} {\bibfnamefont {K.~M.}\ \bibnamefont {Ho}},\ }\href
  {https://doi.org/10.1063/1.5131500} {\bibfield  {journal} {\bibinfo
  {journal} {The Journal of Chemical Physics}\ }\textbf {\bibinfo {volume}
  {151}},\ \bibinfo {pages} {214502} (\bibinfo {year} {2019})}\BibitemShut
  {NoStop}%
\bibitem [{\citenamefont {Laws}\ \emph {et~al.}(2015)\citenamefont {Laws},
  \citenamefont {Miracle},\ and\ \citenamefont {Ferry}}]{RN112}%
  \BibitemOpen
  \bibfield  {author} {\bibinfo {author} {\bibfnamefont {K.~J.}\ \bibnamefont
  {Laws}}, \bibinfo {author} {\bibfnamefont {D.~B.}\ \bibnamefont {Miracle}},\
  and\ \bibinfo {author} {\bibfnamefont {M.}~\bibnamefont {Ferry}},\ }\href
  {https://doi.org/10.1038/ncomms9123} {\bibfield  {journal} {\bibinfo
  {journal} {Nature Communications}\ }\textbf {\bibinfo {volume} {6}},\
  \bibinfo {pages} {8123} (\bibinfo {year} {2015})}\BibitemShut {NoStop}%
\bibitem [{\citenamefont {Kittel}(2005)}]{RN53}%
  \BibitemOpen
  \bibfield  {author} {\bibinfo {author} {\bibfnamefont {C.}~\bibnamefont
  {Kittel}},\ }\href@noop {} {\emph {\bibinfo {title} {Introduction to solid
  state physics}}}\ (\bibinfo  {publisher} {J. Wiley},\ \bibinfo {year}
  {2005})\BibitemShut {NoStop}%
\bibitem [{\citenamefont {Takeuchi}\ and\ \citenamefont {Inoue}(2005)}]{RN54}%
  \BibitemOpen
  \bibfield  {author} {\bibinfo {author} {\bibfnamefont {A.}~\bibnamefont
  {Takeuchi}}\ and\ \bibinfo {author} {\bibfnamefont {A.}~\bibnamefont
  {Inoue}},\ }\href {https://doi.org/10.2320/matertrans.46.2817} {\bibfield
  {journal} {\bibinfo  {journal} {MATERIALS TRANSACTIONS}\ }\textbf {\bibinfo
  {volume} {46}},\ \bibinfo {pages} {2817} (\bibinfo {year}
  {2005})}\BibitemShut {NoStop}%
\bibitem [{\citenamefont {Jacobson}\ and\ \citenamefont
  {Thompson}(2022)}]{RN64}%
  \BibitemOpen
  \bibfield  {author} {\bibinfo {author} {\bibfnamefont {D.~W.}\ \bibnamefont
  {Jacobson}}\ and\ \bibinfo {author} {\bibfnamefont {G.~B.}\ \bibnamefont
  {Thompson}},\ }\href
  {https://doi.org/https://doi.org/10.1016/j.commatsci.2022.111206} {\bibfield
  {journal} {\bibinfo  {journal} {Computational Materials Science}\ }\textbf
  {\bibinfo {volume} {205}},\ \bibinfo {pages} {111206} (\bibinfo {year}
  {2022})}\BibitemShut {NoStop}%
\bibitem [{\citenamefont {Finnis}\ and\ \citenamefont
  {Sinclair}(1984)}]{RN107}%
  \BibitemOpen
  \bibfield  {author} {\bibinfo {author} {\bibfnamefont {M.~W.}\ \bibnamefont
  {Finnis}}\ and\ \bibinfo {author} {\bibfnamefont {J.~E.}\ \bibnamefont
  {Sinclair}},\ }\href {https://doi.org/10.1080/01418618408244210} {\bibfield
  {journal} {\bibinfo  {journal} {Philosophical Magazine A}\ }\textbf {\bibinfo
  {volume} {50}},\ \bibinfo {pages} {45} (\bibinfo {year} {1984})}\BibitemShut
  {NoStop}%
\bibitem [{\citenamefont {Zhang}\ \emph
  {et~al.}(2022{\natexlab{b}})\citenamefont {Zhang}, \citenamefont {Ding},\
  and\ \citenamefont {Ma}}]{RN93}%
  \BibitemOpen
  \bibfield  {author} {\bibinfo {author} {\bibfnamefont {Z.}~\bibnamefont
  {Zhang}}, \bibinfo {author} {\bibfnamefont {J.}~\bibnamefont {Ding}},\ and\
  \bibinfo {author} {\bibfnamefont {E.}~\bibnamefont {Ma}},\ }\href
  {https://doi.org/doi:10.1073/pnas.2213941119} {\bibfield  {journal} {\bibinfo
   {journal} {Proceedings of the National Academy of Sciences}\ }\textbf
  {\bibinfo {volume} {119}},\ \bibinfo {pages} {e2213941119} (\bibinfo {year}
  {2022}{\natexlab{b}})}\BibitemShut {NoStop}%
\bibitem [{\citenamefont {Tang}\ and\ \citenamefont {Harrowell}(2013)}]{RN108}%
  \BibitemOpen
  \bibfield  {author} {\bibinfo {author} {\bibfnamefont {C.}~\bibnamefont
  {Tang}}\ and\ \bibinfo {author} {\bibfnamefont {P.}~\bibnamefont
  {Harrowell}},\ }\href {https://doi.org/10.1038/nmat3631} {\bibfield
  {journal} {\bibinfo  {journal} {Nature Materials}\ }\textbf {\bibinfo
  {volume} {12}},\ \bibinfo {pages} {507} (\bibinfo {year} {2013})}\BibitemShut
  {NoStop}%
\bibitem [{\citenamefont {Şopu}\ \emph {et~al.}(2016)\citenamefont {Şopu},
  \citenamefont {Foroughi}, \citenamefont {Stoica},\ and\ \citenamefont
  {Eckert}}]{RN83}%
  \BibitemOpen
  \bibfield  {author} {\bibinfo {author} {\bibfnamefont {D.}~\bibnamefont
  {Şopu}}, \bibinfo {author} {\bibfnamefont {A.}~\bibnamefont {Foroughi}},
  \bibinfo {author} {\bibfnamefont {M.}~\bibnamefont {Stoica}},\ and\ \bibinfo
  {author} {\bibfnamefont {J.}~\bibnamefont {Eckert}},\ }\href
  {https://doi.org/10.1021/acs.nanolett.6b01636} {\bibfield  {journal}
  {\bibinfo  {journal} {Nano Letters}\ }\textbf {\bibinfo {volume} {16}},\
  \bibinfo {pages} {4467} (\bibinfo {year} {2016})}\BibitemShut {NoStop}%
\bibitem [{\citenamefont {Steinhardt}\ \emph {et~al.}(1983)\citenamefont
  {Steinhardt}, \citenamefont {Nelson},\ and\ \citenamefont
  {Ronchetti}}]{RN61}%
  \BibitemOpen
  \bibfield  {author} {\bibinfo {author} {\bibfnamefont {P.~J.}\ \bibnamefont
  {Steinhardt}}, \bibinfo {author} {\bibfnamefont {D.~R.}\ \bibnamefont
  {Nelson}},\ and\ \bibinfo {author} {\bibfnamefont {M.}~\bibnamefont
  {Ronchetti}},\ }\href {https://doi.org/10.1103/PhysRevB.28.784} {\bibfield
  {journal} {\bibinfo  {journal} {Physical Review B}\ }\textbf {\bibinfo
  {volume} {28}},\ \bibinfo {pages} {784} (\bibinfo {year} {1983})}\BibitemShut
  {NoStop}%
\bibitem [{\citenamefont {Leocmach}\ and\ \citenamefont {Tanaka}(2012)}]{RN85}%
  \BibitemOpen
  \bibfield  {author} {\bibinfo {author} {\bibfnamefont {M.}~\bibnamefont
  {Leocmach}}\ and\ \bibinfo {author} {\bibfnamefont {H.}~\bibnamefont
  {Tanaka}},\ }\href {https://doi.org/10.1038/ncomms1974} {\bibfield  {journal}
  {\bibinfo  {journal} {Nature Communications}\ }\textbf {\bibinfo {volume}
  {3}},\ \bibinfo {pages} {974} (\bibinfo {year} {2012})}\BibitemShut {NoStop}%
\bibitem [{\citenamefont {Hu}\ \emph {et~al.}(2020)\citenamefont {Hu},
  \citenamefont {Zhang}, \citenamefont {Kube}, \citenamefont {Schroers},
  \citenamefont {Shattuck},\ and\ \citenamefont {O'Hern}}]{RN86}%
  \BibitemOpen
  \bibfield  {author} {\bibinfo {author} {\bibfnamefont {Y.-C.}\ \bibnamefont
  {Hu}}, \bibinfo {author} {\bibfnamefont {K.}~\bibnamefont {Zhang}}, \bibinfo
  {author} {\bibfnamefont {S.~A.}\ \bibnamefont {Kube}}, \bibinfo {author}
  {\bibfnamefont {J.}~\bibnamefont {Schroers}}, \bibinfo {author}
  {\bibfnamefont {M.~D.}\ \bibnamefont {Shattuck}},\ and\ \bibinfo {author}
  {\bibfnamefont {C.~S.}\ \bibnamefont {O'Hern}},\ }\href
  {https://doi.org/10.1103/PhysRevMaterials.4.105602} {\bibfield  {journal}
  {\bibinfo  {journal} {Physical Review Materials}\ }\textbf {\bibinfo {volume}
  {4}},\ \bibinfo {pages} {105602} (\bibinfo {year} {2020})}\BibitemShut
  {NoStop}%
\bibitem [{\citenamefont {Rycroft}(2009)}]{RN58}%
  \BibitemOpen
  \bibfield  {author} {\bibinfo {author} {\bibfnamefont {C.~H.}\ \bibnamefont
  {Rycroft}},\ }\href {https://doi.org/10.1063/1.3215722} {\bibfield  {journal}
  {\bibinfo  {journal} {Chaos: An Interdisciplinary Journal of Nonlinear
  Science}\ }\textbf {\bibinfo {volume} {19}},\ \bibinfo {pages} {041111}
  (\bibinfo {year} {2009})}\BibitemShut {NoStop}%
\bibitem [{\citenamefont {Mickel}\ \emph {et~al.}(2013)\citenamefont {Mickel},
  \citenamefont {Kapfer}, \citenamefont {Schröder-Turk},\ and\ \citenamefont
  {Mecke}}]{RN60}%
  \BibitemOpen
  \bibfield  {author} {\bibinfo {author} {\bibfnamefont {W.}~\bibnamefont
  {Mickel}}, \bibinfo {author} {\bibfnamefont {S.~C.}\ \bibnamefont {Kapfer}},
  \bibinfo {author} {\bibfnamefont {G.~E.}\ \bibnamefont {Schröder-Turk}},\
  and\ \bibinfo {author} {\bibfnamefont {K.}~\bibnamefont {Mecke}},\ }\href
  {https://doi.org/10.1063/1.4774084} {\bibfield  {journal} {\bibinfo
  {journal} {The Journal of Chemical Physics}\ }\textbf {\bibinfo {volume}
  {138}},\ \bibinfo {pages} {044501} (\bibinfo {year} {2013})}\BibitemShut
  {NoStop}%
\bibitem [{\citenamefont {Lechner}\ and\ \citenamefont {Dellago}(2008)}]{RN59}%
  \BibitemOpen
  \bibfield  {author} {\bibinfo {author} {\bibfnamefont {W.}~\bibnamefont
  {Lechner}}\ and\ \bibinfo {author} {\bibfnamefont {C.}~\bibnamefont
  {Dellago}},\ }\href {https://doi.org/10.1063/1.2977970} {\bibfield  {journal}
  {\bibinfo  {journal} {The Journal of Chemical Physics}\ }\textbf {\bibinfo
  {volume} {129}},\ \bibinfo {pages} {114707} (\bibinfo {year}
  {2008})}\BibitemShut {NoStop}%
\bibitem [{\citenamefont {Heinz}\ \emph {et~al.}(2005)\citenamefont {Heinz},
  \citenamefont {Paul},\ and\ \citenamefont {Binder}}]{RN14}%
  \BibitemOpen
  \bibfield  {author} {\bibinfo {author} {\bibfnamefont {H.}~\bibnamefont
  {Heinz}}, \bibinfo {author} {\bibfnamefont {W.}~\bibnamefont {Paul}},\ and\
  \bibinfo {author} {\bibfnamefont {K.}~\bibnamefont {Binder}},\ }\href
  {https://doi.org/10.1103/PhysRevE.72.066704} {\bibfield  {journal} {\bibinfo
  {journal} {Physical Review E}\ }\textbf {\bibinfo {volume} {72}},\ \bibinfo
  {pages} {066704} (\bibinfo {year} {2005})}\BibitemShut {NoStop}%
\bibitem [{\citenamefont {Zwanzig}\ \emph {et~al.}(1954)\citenamefont
  {Zwanzig}, \citenamefont {Kirkwood}, \citenamefont {Oppenheim},\ and\
  \citenamefont {Alder}}]{RN63}%
  \BibitemOpen
  \bibfield  {author} {\bibinfo {author} {\bibfnamefont {R.~W.}\ \bibnamefont
  {Zwanzig}}, \bibinfo {author} {\bibfnamefont {J.~G.}\ \bibnamefont
  {Kirkwood}}, \bibinfo {author} {\bibfnamefont {I.}~\bibnamefont
  {Oppenheim}},\ and\ \bibinfo {author} {\bibfnamefont {B.~J.}\ \bibnamefont
  {Alder}},\ }\href {https://doi.org/10.1063/1.1740193} {\bibfield  {journal}
  {\bibinfo  {journal} {The Journal of Chemical Physics}\ }\textbf {\bibinfo
  {volume} {22}},\ \bibinfo {pages} {783} (\bibinfo {year} {1954})}\BibitemShut
  {NoStop}%
\bibitem [{\citenamefont {Debenedetti}\ and\ \citenamefont
  {Stillinger}(2001)}]{RN89}%
  \BibitemOpen
  \bibfield  {author} {\bibinfo {author} {\bibfnamefont {P.~G.}\ \bibnamefont
  {Debenedetti}}\ and\ \bibinfo {author} {\bibfnamefont {F.~H.}\ \bibnamefont
  {Stillinger}},\ }\href {https://doi.org/10.1038/35065704} {\bibfield
  {journal} {\bibinfo  {journal} {Nature}\ }\textbf {\bibinfo {volume} {410}},\
  \bibinfo {pages} {259} (\bibinfo {year} {2001})}\BibitemShut {NoStop}%
\bibitem [{\citenamefont {Kumar}\ \emph {et~al.}(2013)\citenamefont {Kumar},
  \citenamefont {Neibecker}, \citenamefont {Liu},\ and\ \citenamefont
  {Schroers}}]{RN56}%
  \BibitemOpen
  \bibfield  {author} {\bibinfo {author} {\bibfnamefont {G.}~\bibnamefont
  {Kumar}}, \bibinfo {author} {\bibfnamefont {P.}~\bibnamefont {Neibecker}},
  \bibinfo {author} {\bibfnamefont {Y.~H.}\ \bibnamefont {Liu}},\ and\ \bibinfo
  {author} {\bibfnamefont {J.}~\bibnamefont {Schroers}},\ }\href
  {https://doi.org/10.1038/ncomms2546} {\bibfield  {journal} {\bibinfo
  {journal} {Nature Communications}\ }\textbf {\bibinfo {volume} {4}},\
  \bibinfo {pages} {1536} (\bibinfo {year} {2013})}\BibitemShut {NoStop}%
\bibitem [{\citenamefont {Ketkaew}\ \emph {et~al.}(2018)\citenamefont
  {Ketkaew}, \citenamefont {Chen}, \citenamefont {Wang}, \citenamefont {Datye},
  \citenamefont {Fan}, \citenamefont {Pereira}, \citenamefont {Schwarz},
  \citenamefont {Liu}, \citenamefont {Yamada}, \citenamefont {Dmowski},
  \citenamefont {Shattuck}, \citenamefont {O’Hern}, \citenamefont {Egami},
  \citenamefont {Bouchbinder},\ and\ \citenamefont {Schroers}}]{RN87}%
  \BibitemOpen
  \bibfield  {author} {\bibinfo {author} {\bibfnamefont {J.}~\bibnamefont
  {Ketkaew}}, \bibinfo {author} {\bibfnamefont {W.}~\bibnamefont {Chen}},
  \bibinfo {author} {\bibfnamefont {H.}~\bibnamefont {Wang}}, \bibinfo {author}
  {\bibfnamefont {A.}~\bibnamefont {Datye}}, \bibinfo {author} {\bibfnamefont
  {M.}~\bibnamefont {Fan}}, \bibinfo {author} {\bibfnamefont {G.}~\bibnamefont
  {Pereira}}, \bibinfo {author} {\bibfnamefont {U.~D.}\ \bibnamefont
  {Schwarz}}, \bibinfo {author} {\bibfnamefont {Z.}~\bibnamefont {Liu}},
  \bibinfo {author} {\bibfnamefont {R.}~\bibnamefont {Yamada}}, \bibinfo
  {author} {\bibfnamefont {W.}~\bibnamefont {Dmowski}}, \bibinfo {author}
  {\bibfnamefont {M.~D.}\ \bibnamefont {Shattuck}}, \bibinfo {author}
  {\bibfnamefont {C.~S.}\ \bibnamefont {O’Hern}}, \bibinfo {author}
  {\bibfnamefont {T.}~\bibnamefont {Egami}}, \bibinfo {author} {\bibfnamefont
  {E.}~\bibnamefont {Bouchbinder}},\ and\ \bibinfo {author} {\bibfnamefont
  {J.}~\bibnamefont {Schroers}},\ }\href
  {https://doi.org/10.1038/s41467-018-05682-8} {\bibfield  {journal} {\bibinfo
  {journal} {Nature Communications}\ }\textbf {\bibinfo {volume} {9}},\
  \bibinfo {pages} {3271} (\bibinfo {year} {2018})}\BibitemShut {NoStop}%
\bibitem [{\citenamefont {Fan}\ \emph {et~al.}(2017)\citenamefont {Fan},
  \citenamefont {Wang}, \citenamefont {Zhang}, \citenamefont {Liu},
  \citenamefont {Schroers}, \citenamefont {Shattuck},\ and\ \citenamefont
  {O'Hern}}]{RN55}%
  \BibitemOpen
  \bibfield  {author} {\bibinfo {author} {\bibfnamefont {M.}~\bibnamefont
  {Fan}}, \bibinfo {author} {\bibfnamefont {M.}~\bibnamefont {Wang}}, \bibinfo
  {author} {\bibfnamefont {K.}~\bibnamefont {Zhang}}, \bibinfo {author}
  {\bibfnamefont {Y.}~\bibnamefont {Liu}}, \bibinfo {author} {\bibfnamefont
  {J.}~\bibnamefont {Schroers}}, \bibinfo {author} {\bibfnamefont {M.~D.}\
  \bibnamefont {Shattuck}},\ and\ \bibinfo {author} {\bibfnamefont {C.~S.}\
  \bibnamefont {O'Hern}},\ }\href {https://doi.org/10.1103/PhysRevE.95.022611}
  {\bibfield  {journal} {\bibinfo  {journal} {Physical Review E}\ }\textbf
  {\bibinfo {volume} {95}},\ \bibinfo {pages} {022611} (\bibinfo {year}
  {2017})}\BibitemShut {NoStop}%
\bibitem [{\citenamefont {Becker}\ \emph {et~al.}(2013)\citenamefont {Becker},
  \citenamefont {Tavazza}, \citenamefont {Trautt},\ and\ \citenamefont
  {Buarque~de Macedo}}]{RN113}%
  \BibitemOpen
  \bibfield  {author} {\bibinfo {author} {\bibfnamefont {C.~A.}\ \bibnamefont
  {Becker}}, \bibinfo {author} {\bibfnamefont {F.}~\bibnamefont {Tavazza}},
  \bibinfo {author} {\bibfnamefont {Z.~T.}\ \bibnamefont {Trautt}},\ and\
  \bibinfo {author} {\bibfnamefont {R.~A.}\ \bibnamefont {Buarque~de Macedo}},\
  }\href {https://doi.org/https://doi.org/10.1016/j.cossms.2013.10.001}
  {\bibfield  {journal} {\bibinfo  {journal} {Current Opinion in Solid State
  and Materials Science}\ }\textbf {\bibinfo {volume} {17}},\ \bibinfo {pages}
  {277} (\bibinfo {year} {2013})}\BibitemShut {NoStop}%
\bibitem [{\citenamefont {Hale}\ \emph {et~al.}(2018)\citenamefont {Hale},
  \citenamefont {Trautt},\ and\ \citenamefont {Becker}}]{RN114}%
  \BibitemOpen
  \bibfield  {author} {\bibinfo {author} {\bibfnamefont {L.~M.}\ \bibnamefont
  {Hale}}, \bibinfo {author} {\bibfnamefont {Z.~T.}\ \bibnamefont {Trautt}},\
  and\ \bibinfo {author} {\bibfnamefont {C.~A.}\ \bibnamefont {Becker}},\
  }\href {https://doi.org/10.1088/1361-651X/aabc05} {\bibfield  {journal}
  {\bibinfo  {journal} {Modelling and Simulation in Materials Science and
  Engineering}\ }\textbf {\bibinfo {volume} {26}},\ \bibinfo {pages} {055003}
  (\bibinfo {year} {2018})}\BibitemShut {NoStop}%
\bibitem [{\citenamefont {{NIST Interatomic Potentials Repository}}()}]{RN115}%
  \BibitemOpen
  \bibfield  {author} {\bibinfo {author} {\bibnamefont {{NIST Interatomic
  Potentials Repository}}},\ }\href@noop {} {}\bibinfo {note}
  {\url{https://www.ctcms.nist.gov/potentials}, Last accessed on
  01-17-2023}\BibitemShut {NoStop}%
\bibitem [{\citenamefont {Liu}\ \emph {et~al.}(2021{\natexlab{b}})\citenamefont
  {Liu}, \citenamefont {Ma}, \citenamefont {Liao}, \citenamefont {Liu},
  \citenamefont {Mota}, \citenamefont {Liu}, \citenamefont {Sohn},
  \citenamefont {Kube}, \citenamefont {Zhao}, \citenamefont {Singer},\ and\
  \citenamefont {Schroers}}]{RN47}%
  \BibitemOpen
  \bibfield  {author} {\bibinfo {author} {\bibfnamefont {N.}~\bibnamefont
  {Liu}}, \bibinfo {author} {\bibfnamefont {T.}~\bibnamefont {Ma}}, \bibinfo
  {author} {\bibfnamefont {C.}~\bibnamefont {Liao}}, \bibinfo {author}
  {\bibfnamefont {G.}~\bibnamefont {Liu}}, \bibinfo {author} {\bibfnamefont
  {R.~M.~O.}\ \bibnamefont {Mota}}, \bibinfo {author} {\bibfnamefont
  {J.}~\bibnamefont {Liu}}, \bibinfo {author} {\bibfnamefont {S.}~\bibnamefont
  {Sohn}}, \bibinfo {author} {\bibfnamefont {S.}~\bibnamefont {Kube}}, \bibinfo
  {author} {\bibfnamefont {S.}~\bibnamefont {Zhao}}, \bibinfo {author}
  {\bibfnamefont {J.~P.}\ \bibnamefont {Singer}},\ and\ \bibinfo {author}
  {\bibfnamefont {J.}~\bibnamefont {Schroers}},\ }\href
  {https://doi.org/10.1038/s41598-021-83384-w} {\bibfield  {journal} {\bibinfo
  {journal} {Scientific Reports}\ }\textbf {\bibinfo {volume} {11}},\ \bibinfo
  {pages} {3903} (\bibinfo {year} {2021}{\natexlab{b}})}\BibitemShut {NoStop}%
\bibitem [{\citenamefont {Bordeenithikasem}\ \emph {et~al.}(2017)\citenamefont
  {Bordeenithikasem}, \citenamefont {Liu}, \citenamefont {Kube}, \citenamefont
  {Li}, \citenamefont {Ma}, \citenamefont {Scanley}, \citenamefont
  {Broadbridge}, \citenamefont {Vlassak}, \citenamefont {Singer},\ and\
  \citenamefont {Schroers}}]{RN48}%
  \BibitemOpen
  \bibfield  {author} {\bibinfo {author} {\bibfnamefont {P.}~\bibnamefont
  {Bordeenithikasem}}, \bibinfo {author} {\bibfnamefont {J.}~\bibnamefont
  {Liu}}, \bibinfo {author} {\bibfnamefont {S.~A.}\ \bibnamefont {Kube}},
  \bibinfo {author} {\bibfnamefont {Y.}~\bibnamefont {Li}}, \bibinfo {author}
  {\bibfnamefont {T.}~\bibnamefont {Ma}}, \bibinfo {author} {\bibfnamefont
  {B.~E.}\ \bibnamefont {Scanley}}, \bibinfo {author} {\bibfnamefont {C.~C.}\
  \bibnamefont {Broadbridge}}, \bibinfo {author} {\bibfnamefont {J.~J.}\
  \bibnamefont {Vlassak}}, \bibinfo {author} {\bibfnamefont {J.~P.}\
  \bibnamefont {Singer}},\ and\ \bibinfo {author} {\bibfnamefont
  {J.}~\bibnamefont {Schroers}},\ }\href
  {https://doi.org/10.1038/s41598-017-07719-2} {\bibfield  {journal} {\bibinfo
  {journal} {Scientific Reports}\ }\textbf {\bibinfo {volume} {7}},\ \bibinfo
  {pages} {7155} (\bibinfo {year} {2017})}\BibitemShut {NoStop}%
\bibitem [{\citenamefont {Kube}\ \emph {et~al.}(2019)\citenamefont {Kube},
  \citenamefont {Sohn}, \citenamefont {Uhl}, \citenamefont {Datye},
  \citenamefont {Mehta},\ and\ \citenamefont {Schroers}}]{RN49}%
  \BibitemOpen
  \bibfield  {author} {\bibinfo {author} {\bibfnamefont {S.~A.}\ \bibnamefont
  {Kube}}, \bibinfo {author} {\bibfnamefont {S.}~\bibnamefont {Sohn}}, \bibinfo
  {author} {\bibfnamefont {D.}~\bibnamefont {Uhl}}, \bibinfo {author}
  {\bibfnamefont {A.}~\bibnamefont {Datye}}, \bibinfo {author} {\bibfnamefont
  {A.}~\bibnamefont {Mehta}},\ and\ \bibinfo {author} {\bibfnamefont
  {J.}~\bibnamefont {Schroers}},\ }\href
  {https://doi.org/https://doi.org/10.1016/j.actamat.2019.01.023} {\bibfield
  {journal} {\bibinfo  {journal} {Acta Materialia}\ }\textbf {\bibinfo {volume}
  {166}},\ \bibinfo {pages} {677} (\bibinfo {year} {2019})}\BibitemShut
  {NoStop}%
\bibitem [{\citenamefont {Sato}\ \emph {et~al.}(2005)\citenamefont {Sato},
  \citenamefont {Sanada}, \citenamefont {Saida}, \citenamefont {Imafuku},
  \citenamefont {Matsubara},\ and\ \citenamefont {Inoue}}]{RN45}%
  \BibitemOpen
  \bibfield  {author} {\bibinfo {author} {\bibfnamefont {S.}~\bibnamefont
  {Sato}}, \bibinfo {author} {\bibfnamefont {T.}~\bibnamefont {Sanada}},
  \bibinfo {author} {\bibfnamefont {J.}~\bibnamefont {Saida}}, \bibinfo
  {author} {\bibfnamefont {M.}~\bibnamefont {Imafuku}}, \bibinfo {author}
  {\bibfnamefont {E.}~\bibnamefont {Matsubara}},\ and\ \bibinfo {author}
  {\bibfnamefont {A.}~\bibnamefont {Inoue}},\ }\href
  {https://doi.org/10.2320/matertrans.46.2893} {\bibfield  {journal} {\bibinfo
  {journal} {MATERIALS TRANSACTIONS}\ }\textbf {\bibinfo {volume} {46}},\
  \bibinfo {pages} {2893} (\bibinfo {year} {2005})}\BibitemShut {NoStop}%
\bibitem [{\citenamefont {Li}\ \emph {et~al.}(2010)\citenamefont {Li},
  \citenamefont {Zhang}, \citenamefont {Dong}, \citenamefont {Qiang},
  \citenamefont {Makino},\ and\ \citenamefont {Inoue}}]{RN46}%
  \BibitemOpen
  \bibfield  {author} {\bibinfo {author} {\bibfnamefont {Y.~H.}\ \bibnamefont
  {Li}}, \bibinfo {author} {\bibfnamefont {W.}~\bibnamefont {Zhang}}, \bibinfo
  {author} {\bibfnamefont {C.}~\bibnamefont {Dong}}, \bibinfo {author}
  {\bibfnamefont {J.~B.}\ \bibnamefont {Qiang}}, \bibinfo {author}
  {\bibfnamefont {A.}~\bibnamefont {Makino}},\ and\ \bibinfo {author}
  {\bibfnamefont {A.}~\bibnamefont {Inoue}},\ }\href
  {https://doi.org/https://doi.org/10.1016/j.intermet.2010.03.041} {\bibfield
  {journal} {\bibinfo  {journal} {Intermetallics}\ }\textbf {\bibinfo {volume}
  {18}},\ \bibinfo {pages} {1851} (\bibinfo {year} {2010})}\BibitemShut
  {NoStop}%
\bibitem [{\citenamefont {Ozawa}\ \emph {et~al.}(2018)\citenamefont {Ozawa},
  \citenamefont {Berthier}, \citenamefont {Biroli}, \citenamefont {Rosso},\
  and\ \citenamefont {Tarjus}}]{RN110}%
  \BibitemOpen
  \bibfield  {author} {\bibinfo {author} {\bibfnamefont {M.}~\bibnamefont
  {Ozawa}}, \bibinfo {author} {\bibfnamefont {L.}~\bibnamefont {Berthier}},
  \bibinfo {author} {\bibfnamefont {G.}~\bibnamefont {Biroli}}, \bibinfo
  {author} {\bibfnamefont {A.}~\bibnamefont {Rosso}},\ and\ \bibinfo {author}
  {\bibfnamefont {G.}~\bibnamefont {Tarjus}},\ }\href
  {https://doi.org/10.1073/pnas.1806156115} {\bibfield  {journal} {\bibinfo
  {journal} {Proceedings of the National Academy of Sciences}\ }\textbf
  {\bibinfo {volume} {115}},\ \bibinfo {pages} {6656} (\bibinfo {year}
  {2018})}\BibitemShut {NoStop}%
\end{thebibliography}%
\end{document}